\documentclass[11pt]{article}

\makeatletter
\@ifundefined{pdfoutput}{
    \usepackage[dvips]{graphicx}
}{
    \ifnum\pdfoutput=0\relax
    \usepackage[dvips]{graphicx}
    \fi
    \ifnum\pdfoutput=1\relax
    \usepackage[pdftex]{graphicx}
    \fi
}
\makeatother

\usepackage{url}
\usepackage{latexsym}
\usepackage{amsfonts}
\usepackage{amsmath}
\usepackage{amssymb}
\usepackage{color}
\usepackage{a4wide}
\usepackage{todonotes}
\usepackage{authblk}
\usepackage{booktabs}
\usepackage{cprotect}

\makeatletter
\def\mdseries@tt{m}
\makeatother
\usepackage[frozencache]{minted}

\usepackage{pifont}
\newcommand{\cmark}{\ding{51}}%
\newcommand{\xmark}{\ding{55}}%

\usepackage{siunitx}
\usepackage{tikz} 
\usepackage{pgfplots}

\pgfplotsset{compat=newest} 
\usepgfplotslibrary{units, groupplots} 

\newcommand{\jgrapht}{{J}{G}raph{T}}
\newcommand{\jgraphtS}{{J}{G}raph{T} {S}parse}
\newcommand{\igraph}{igraph}
\newcommand{\bgl}{{B}{G}{L}}

\newcommand{\jung}{{J}ung}
\newcommand{\nx}{{N}etwork{X}}
\newcommand{\lemon}{{L}emon}
\newcommand{\blossom}{{B}lossom{V}}
\newcommand{\cplex}{{C}{P}{L}{E}{X}}
\newcommand{\guava}{{G}uava}
\newcommand{\guavaN}{{G}uava {N}etwork}

\newcommand{\guavaV}{{G}uava {V}alue{G}raph}
\newcommand{\fastutil}{{f}astutil}
\newcommand{\Oh}{\mathcal{O}}
\newcommand{\cpp}{C$\mathrel{++}$}

\newcommand{\jmint}[1]{\mintinline{Java}{#1}}

\author[1]{Dimitrios Michail}
\affil[1]{Dept. of Informatics and Telematics\authorcr Harokopio University of Athens, Greece\authorcr  michail@hua.gr}
\author[2,3]{Joris Kinable}
\affil[2]{Dept. of Industrial Engineering and Innovation Sciences\authorcr
Eindhoven University of Technology, The Netherlands \authorcr j.kinable@tue.nl}
\affil[3]{Robotics Institute\authorcr
Carnegie Mellon University, Pittsburgh, USA}
\author[4]{Barak Naveh}
\affil[4]{barak\_pub@3pq.com}
\author[5]{John V Sichi}
\affil[5]{The JGraphT project\authorcr jsichi@gmail.com}

\date{}

\begin{document}

\title{\jgrapht{} - A Java library for graph data structures and algorithms}

\maketitle

\begin{abstract}
Mathematical software and graph-theoretical algorithmic packages to efficiently model, analyze
and query graphs are crucial in an era where large-scale spatial, societal and economic network
data are abundantly available. One such package is \jgrapht{}, a programming library which contains
very efficient and generic graph data-structures along with a large collection of state-of-the-art
algorithms. The library is written in Java with stability, interoperability and performance in mind.
A distinctive feature of this library is its ability to model vertices and edges as arbitrary objects,
thereby permitting natural representations of many common networks including transportation, social
and biological networks. Besides classic graph algorithms such as shortest-paths and spanning-tree
algorithms, the library contains numerous advanced algorithms: graph and subgraph isomorphism; matching
and flow problems; approximation algorithms for NP-hard problems such as independent set and TSP; and
several more exotic algorithms such as Berge graph detection. Due to its versatility and generic
design, \jgrapht{} is currently used in large-scale commercial products, as well as non-commercial and academic research
projects.\\
In this work we describe in detail the design and underlying structure of the library, and discuss its
most important features and algorithms. A computational study is conducted to evaluate the performance
of \jgrapht{} versus a number of similar libraries. Experiments on a large number of graphs over a
variety of popular algorithms show that \jgrapht{} is highly competitive with other established libraries
such as \nx{} or the \bgl{}.
\end{abstract}

\section{Introduction}

Over the last decade, a surge in demand for studying large, complex graphs spurred the development of new packages for
graph analysis. Graphs became ubiquitous in every field of study due to their natural ability to capture relationships
and interactions between different entities. Graph theoretical problems are regularly encountered in such diverse areas
as network security, computational biology, logistics/planning, psychology, chemistry, and linguistics. Despite the vast
diversity in graph applications across different fields, their underlying mechanics inevitably rely on the same fundamental
mathematical techniques and solution approaches. In keeping with this observation, libraries which efficiently model, store,
manipulate and query graphs have become indispensable for engineers and data scientists alike.

This paper introduces \jgrapht{}, a library which contains very efficient and generic graph data-structures along with a
sizeable collection of sophisticated algorithms. The library is written in Java, with stability, performance and
interoperability in mind. The first version of \jgrapht{}, released in 2003, was primarily intended as a scientific package
containing graph-theoretical algorithms. Over the years, \jgrapht{} widened its scope, and added support for algorithms typically
encountered in the context of (path) planning, routing, network analysis, combinatorial optimization and applications in
computational biology. These developments lead to the adoption of \jgrapht{} into large-scale projects both in academia and
industry. As of today, \jgrapht{} is used\footnote{\url{https://mvnrepository.com/artifact/org.jgrapht/jgrapht-core/usages}}
in a variety of commercial and open-source software packages, including the Apache Cassandra database, the distributed realtime
computation system Apache Storm, the Graal JVM, the Constraint Programming Solver Choco, and in Cascading, a software abstraction
layer for Apache Hadoop. Similarly, in academia \jgrapht{} has been successfully deployed across a wide range of research domains,
including circuit verification~\cite{LYRB11}, malware detection~\cite{KIOR11}, software performance prediction~\cite{KROG12},
cartography~\cite{MARS14}, social networking~\cite{BADE11}, and navigation of autonomous vehicles~\cite{DABL05}.

Developing a robust, performance-driven, application-independent graph library is a complex task, involving a large number of
conflicting (functional and structural) design choices and performance trade-offs. In this paper, we formally outline the
design of \jgrapht{}, and highlight several of its design considerations. Moreover, we provide an overview of the most important
features and algorithms currently supported by \jgrapht{}. Among others, this overview covers routing algorithms such as shortest
path algorithms or advanced heuristics for A*; network analysis with clustering coefficients and centrality metrics; network
optimization and matching problems; min-cut and max-flow algorithms; graph mining with graph kernels; and subgraph isomorphism
detection. To show \jgrapht{}'s competitiveness, we perform a computational comparison with other well-established graph libraries.

The remainder of this paper is structured as follows. Section~\ref{sec:related} discusses related work and alternative graph
libraries. Next, Section~\ref{sec:design} describes \jgrapht{}, its components and its internal design in detail. An overview
of the various algorithms supported by \jgrapht{} is provided in Section~\ref{sec:algorithms}, followed by an overview of
graph generators in Section~\ref{sec:generators}. To provide interoperability between different mathematical packages,
\jgrapht{} natively supports a large variety of graph formats, summarized in Section~\ref{sec:io}. An extensive computational
study---covering an external comparison of algorithms from different libraries, an internal comparison of alternative algorithms
for the same mathematical problems, and a comparison of different graph
representations---is presented in Section~\ref{sec:evaluation}.
Finally, Section~\ref{sec:conclusions} concludes the paper.

\section{Related Work} \label{sec:related}

Software solutions for graph theory exist in many forms. On one side of the spectrum, there are the mathematical ecosystems
such as Wolfram Mathematica~\cite{MATH18}, Sage Math~\cite{SAGE18} and Maple~\cite{MAPL18} which provide high-level functions
to model, analyze and visualize graphs and networks. On the other side, there are graph theoretical libraries and algorithmic
packages such as \jgrapht{}, which are primarily designed to aid software development. From a scientific point of view, the
two best-known libraries are LEDA~\cite{mehlhorn1995leda} and the Boost Graph Library~\cite{siek2001boost} (BGL), which are
both written in \cpp{}. LEDA offers a very efficient graph data-structure, along with some of the most efficient implementations
of classic graph algorithms. BGL, on the other hand, follows a generic programming paradigm in order to provide highly optimized
graph algorithms. 

Despite the popularity of Java as a programming language in academia and industry, the number of graph packages written in Java
is very limited. Currently, there exist only two viable alternatives to \jgrapht{}: the Java Universal Network/Graph Framework
(JUNG) library~\cite{o2003jung} and a graph component in the Google Guava\footnote{\url{https://github.com/google/guava}} library.
JUNG provides a graph data-structure, several basic algorithms such as shortest paths and centrality metrics, and a graph drawing
(layout) component. Google Guava, on the other hand, currently only contains a number of graph data structures, including 'Graph',
'ValueGraph' and 'Network'. Out of these three, 'Network' is the most general one and corresponds almost one-to-one with the \jgrapht{}
graph interface. To provide algorithmic support for Guava, \jgrapht{} contains adapter classes which allow users to invoke all
algorithms in  \jgrapht{} on Guava graph data-structures.

Software packages for network analysis can be broadly categorized as: (1) packages for data structures and storage, including
databases for large-scale networks; (2) algorithmic packages for network analysis, primarily meant to create insights into the
network data; and (3) packages for graph visualizations to generate meaningful, human-interpretable, visual representations.
Of particular interest to us are the packages that fall within the second category.
The igraph~\cite{csardi2006igraph} library, written in C, contains several optimized algorithms for network analysis. igraph
is designed to handle large graphs efficiently and to be easily embeddable in higher level programming languages such as Python
and R. NetworkX~\cite{hagberg2008exploring} is a Python library designed to study the structure and dynamics of complex networks.
It contains data structures for graphs, digraphs and multigraphs, as well as many standard graph and network analysis
algorithms. Moreover, similar to \jgrapht{}, \nx{} is platform independent.
NetworKit~\cite{staudt2016networkit} is yet another open-source package for large-scale network analysis. It is written in C++,
employing parallelization when appropriate, and provides Python bindings for ease-of-use. The Stanford Network Analysis Platform
(SNAP)~\cite{leskovec2016snap} is a general-purpose, high-performance system that provides easy-to-use, high-level operations for
analysis and manipulation of large networks. The library focuses on single big-memory machines and provides a large collection of
graph algorithms including dynamic algorithms. Similarly to the other libraries, it is written in C++ with Python bindings.

Next to the traditional graph libraries, there exist a number of specialized libraries designed for large-scale,
parallel computing applications. These libraries typically implement frameworks that rely on distributed computing
(Parallel Boost Graph Library~\cite{GRLU05}, Distributed GraphLab~\cite{LBGG12}),
multi-core CPU (Ligra~\cite{SHBL13}, GraphMat~\cite{SSPD15}), or GPU architectures
(GraphBLAST~\cite{YABO19}, Gunrock~\cite{WDPW16}, nvgraph\footnote{\url{https://developer.nvidia.com/nvgraph}})
to execute graph operations on massive graphs in parallel.
Several of these frameworks,
including GraphMat, GraphBLAS~\cite{kepner2016mathematical}\footnote{\url{http://graphblas.org/}} and GraphBLAST,
represent graphs through
sparse adjancency matrices, and use matrix algebra to implement graph operations. 
These frameworks are, among others,
particularly suited for implementing parallel graph traversal algorithms such as Breadth-First-Search, Pagerank
and single-source shortest paths. Empirical studies and comparisons of a number of these libraries have been
published in~\cite{satish2014navigating, GBVI14}. Additionally, a community effort has been launched~\cite{mattson2019lagraph} 
in order to implement more graph algorithms using the GraphBLAS API.

Finally, there exist a large number of software packages and libraries that focus on graph visualization such as
Gephi~\cite{bastian2009gephi}, Cytoscape~\cite{SMOB03}, and GraphViz~\cite{ellson2001graphviz}.
While it is possible to couple \jgrapht{} with visualization libraries, the library currently does not offer drawing
capabilities by itself.

\section{Design}\label{sec:design}

\jgrapht{} is designed with a strong focus on flexibility, versatility and performance. This section outlines the design of
\jgrapht{} and discusses trade-offs and considerations encountered in the design of the library.

\subsection{The Graph interface}

\jgrapht{} is built around a central \jmint{Graph<V,E>} interface (Figure~\ref{fig:jgraphtuml}). This interface provides elementary
operations for the construction of a graph, as well as basic operators to access elements of the graph (Figure~\ref{fig:maininterface}).
All interactions with the graph occur through this interface: every predefined graph class in the library implements this interface,
and all of \jgrapht{}'s algorithms expect a \jmint{Graph} instance as input.

The interface takes two generic parameters \jmint{<V>} and \jmint{<E>} determining the type of Java objects that are used respectively
as vertices and edges of the graph. \jgrapht{} permits the user to use any type of object as edge or vertex. In its simplest form, the
vertices of a graph are represented by Integers or Strings, while the edges are represented by a default edge implementation called
\jmint{DefaultEdge}. A more meaningful example arises when modeling a road network as a graph, where the vertices are intersections
and the edges are road segments. Typically, one would implement an \jmint{Intersection} class which stores the geographical coordinates
of an intersection, as well as a \jmint{RoadSegment} class which records information such as number of lanes, driving speed, length,
shape and perhaps the name of the segment. The possibility to use any type of object as a vertex or edge makes \jgrapht{} extremely
versatile, as its basic data structures are capable of capturing and expressing any type of relationship or interaction between any
type of object in a natural way.

\jgrapht{} provides implementations of common graph types such as simple graphs, multigraphs, pseudographs, etc. Each of these
graph types can be refined as directed or undirected, and weighted or unweighted. An overview of predefined graph types can be
found in Table~\ref{table:graphclasses}. Since each graph implements the aforementioned \jmint{Graph} interface, several methods
behave differently depending on the type of graph. The method \jmint{degreeOf(V vertex)}, for instance, returns the number of edges
touching a vertex (with self-loops counted twice) in case of an undirected graph, whereas the same method returns the sum of the
in-degree and the out-degree in case of a directed graph. Similarly, the \jmint{inDegreeOf(V vertex)} method in a directed graph
returns the number of directed edges leaving the vertex while for  undirected graphs it returns the number of edges touching the
vertex.

To create a new instance of, for example, a simple graph, a user can invoke:
\begin{minted}{java}
Graph<Integer, DefaultEdge> graph = new SimpleGraph<>(DefaultEdge.class);
\end{minted}
Choosing a particular graph implementation, however, can be non-trivial for users foreign to graph theoretical concepts. One
potential strategy to circumvent this issue is to select the most general graph implementation by default. For instance, a
pseudograph which supports multiple edges and self-loops can be used to represent a simple graph which does not support these
features. This however comes with a clear performance penalty, since pseudographs typically take more space, and operations on
these graphs take more time than their more specialized counterparts. To circumvent this issue, and to simplify the process of
selecting the desired type of graph, \jgrapht{} allows the user to construct graphs through a builder pattern~\cite{gamma1993design}
after which the library automatically determines the most suitable graph implementation: 

\begin{minted}{Java}
Graph<Integer, DefaultEdge> graph = 
    GraphTypeBuilder.<Integer, DefaultEdge> directed()
                    .allowingMultipleEdges(true)
                    .allowingSelfLoops(false).
                    .edgeClass(DefaultEdge.class)
                    .buildGraph();
\end{minted}

Algorithms which behave differently depending on the underlying graph characteristics, can query the graph during 
runtime for its \mintinline{Java}{GraphType}. The \mintinline{Java}{GraphType} contains the necessary type information, defining
whether the graph is directed or undirected, weighted or unweighted, and whether it allows self-loops, multiple-edges, etc.

\subsection{Graph structure} \label{subsec:graphstructure}

The structural design of \jgrapht{}, as depicted in Figure~\ref{fig:jgraphtuml}, separates the functional \jmint{Graph<V,E>} interface, from the underlying data structures used to store the graph. The Graph interface, at the top of the hierarchy, defines all high-level graph operations. The class \mintinline{Java}{AbstractGraph} offers a minimal implementation of this Graph interface, without explicitly defining the data structures for storage and indexing as these are managed by the graph \emph{backend}. The graph backend extends \mintinline{Java}{AbstractGraph}, records the \mintinline{Java}{GraphType}, and implements data structures to physically store the graph. Figure~\ref{fig:jgraphtuml} depicts the \emph{default} backend, implemented by the \mintinline{Java}{AbstractBaseGraph} class. Most of the predefined graph classes listed in Table~\ref{table:graphclasses} are subclasses of the \mintinline{Java}{AbstractBaseGraph} class. A detailed discussion of the different graph backends supported by \jgrapht{} is provided in Section \ref{subsec:representations}.

To implement a new graph type, or to adjust the underlying implementation of an existing graph type, the user would typically
instantiate, override or extend some of the classes depicted in Figure~\ref{fig:jgraphtuml}. {\em Views} over graphs, for instance, can be defined by extending \mintinline{Java}{AbstractGraph}. This is similar to the concept of filtered graphs in \bgl{}. All operations invoked on a view are delegated to the graph backing the view. Consequently, views offer a natural way to model, for instance, induced subgraphs (\mintinline{Java}{AsSubgraph}). They can
also be used to treat a directed graph as an undirected graph (\mintinline{Java}{AsUndirectedGraph}), to add weights to an
unweighted graph (\mintinline{Java}{AsWeightedGraph}) or to render a graph unmodifiable (\mintinline{Java}{AsUnModifiableGraph}).
In addition to views, it is possible to define \textit{adapter} classes by extending \mintinline{Java}{AbstractGraph}. One
such example can be found in the \verb|jgrapht-guava| package, which implements adapters for graph-data structures encountered
in the Guava Library\footnote{\url{https://github.com/google/guava}}. Through these adapters, a user can invoke all algorithms
described in Section~\ref{sec:algorithms} on graphs implemented with Guava data-structures.

\subsection{Graph backends}\label{subsec:representations}

The underlying implementation and data storage of a graph, independent of whether the graph type is predefined or user-defined, is highly customizable. There exist many scenarios where domain specific knowledge of the end-user is
required to determine the best choice of data structures. Particularly relevant in this context are: type of data being represented; graph density (sparse or dense graph); graph size (number of edges/vertices); available storage space; performance requirements; and the type of graph operations that will be most frequently performed. Similar considerations are made when explicitly storing an \textit{adjacency matrix} to lookup adjacent vertices (neighbors) or when selecting the structures used to represent the \textit{incidence matrix}. If for instance the vertices are simple integers, the incidence matrix can be a 2-dimensional array, whereas in case of arbitrary vertex objects we must resort to hash tables.

\begin{figure}[t]
\centering
\includegraphics[width=.8\linewidth]{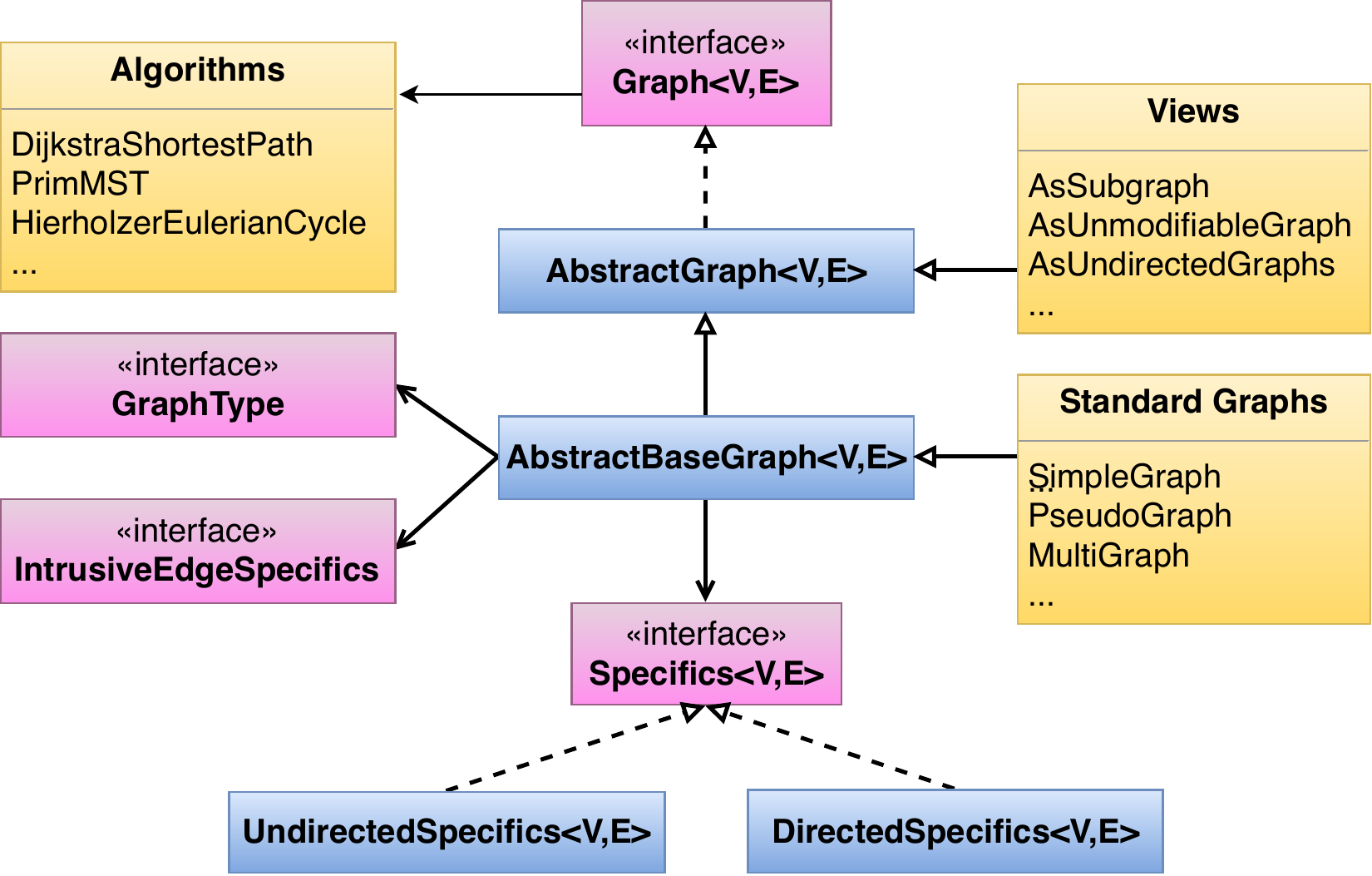}
\caption{Core structure of \jgrapht{}}
\label{fig:jgraphtuml}
\end{figure}

Fine-grained control over how the data in a graph is stored can be obtained by adjusting the graph \emph{backend}. \jgrapht{} provides two predefined backends which collectively cover most common use-cases: the \emph{default} backend and the \emph{sparse} backend. In addition, the user can implement custom backends, simply by creating a new subclass of the \mintinline{Java}{AbstractGraph} class (Figure~\ref{fig:jgraphtuml}). It is worth noting that when using the predefined backends, all operations invoked on \jgrapht{} graphs are performed in a deterministic fashion.
This behavior is realized through the usage
of data structures having a predictable iteration order such as lists, as well as sets and maps backed by doubly linked
lists (\mintinline{Java}{LinkedHashSet} and \mintinline{Java}{LinkedHashMap}).

The \emph{default} backend, implemented by the \mintinline{Java}{AbstractBaseGraph} class (Figure~\ref{fig:jgraphtuml}), is designed to offer a good trade-off between performance and memory consumption. Since vertices and edges can be modeled by arbitrary objects, the default backend primarily relies on hash tables to store vertices and edges, and to implement adjacency lists. Consequently, basic operations such as vertex or edge removal and addition can be performed in expected constant ($\Oh(1)$) time. An (optional) indexing mechanism is provided to index edges by their two endpoints to enable fast edge lookups. This indexing mechanism again provides a trade-off between performance and additional memory consumption. The user can further customize how the adjacency and incidence matrices are stored and how edge lookups are performed (including how edge weights are stored) by providing alternative implementations of the \mintinline{Java}{Specifics} and \mintinline{Java}{IntrusiveEdgesSpecifics} interfaces. For instance, the standard Java hash tables used to store the adjacency and incidence matrices can be swapped out by specialized alternatives from the Fastutil\footnote{\url{https://github.com/vigna/fastutil}} library or from the 
Eclipse Collections\footnote{\url{https://github.com/eclipse/eclipse-collections}} to reduce the memory footprint of a graph.

The default backend is well-suited for general-purpose graphs which efficiently support edit
operations such as the addition or removal of a single vertex or edge. Better performance,
however, can be achieved in a less dynamic setting where the graph is constructed in a single
bulk operation. Such a {\em write-once read-many} strategy is very common when executing complex
algorithms on graphs, which usually involve loading a graph from an external source into memory,
executing the algorithms, and querying the final result. To accommodate such a use-case,
\jgrapht{} provides a specialized \emph{sparse} graph backend which implements the
\mintinline{Java}{Graph<Integer,Integer>} interface and hence requires that all vertices and edges
are represented by integers. Here it is assumed that the vertices are numbered $0, \ldots, n-1$
while edges are numbered $0, \ldots, m-1$, where $n$ and $m$ are resp.\ the number of vertices and
edges in the graph. The latter assumption renders edit operations less efficient, since adding or
removing a vertex (resp.\ edge) potentially involves re-numbering all other vertices and edges.
To reduce the storage space requirements of a graph, the sparse backend stores the incidence
matrix in Compressed-Sparse-Rows (CSR) format~\cite{saad2003iterative}, thereby taking advantage of
the fact that most real-world graphs are sparse graphs. Graphs stored in this format support both
self-loops and multiple-edges. An overview of the sparse graphs and their capabilities is provided
in Table~\ref{table:graphclasses}. To implement undirected graphs, the sparse backend represents
the incidence matrix through a single boolean matrix, whereas two boolean matrices (one for each
direction) are used for directed graphs.
Edge weights, in case of weighted graphs, are stored in a simple array of length $m$.

\cprotEnv\begin{figure}
\begin{small}
\begin{minted}{Java}
public interface Graph<V,E> {
    GraphType getType();

    V addVertex();
    boolean removeVertex(V v);
    E addEdge(V sourceVertex, V targetVertex);
    boolean removeEdge(E e);

    Set<V> vertexSet();
    Set<E> edgeSet();
    
    V getEdgeSource(E e);
    V getEdgeTarget(E e);
    E getEdge(V sourceVertex, V targetVertex);
    
    double getEdgeWeight(E e);
    void setEdgeWeight(E e, double weight);
    
    Set<E> edgesOf(V v);

    /* More methods omitted */
}
\end{minted}
\end{small}
\cprotect\caption{The \verb|Graph<V,E>| interface. All interactions with the graph happen through this interface.}
\label{fig:maininterface}
\end{figure}

\begin{table}[t]
  \begin{tabular}{l|p{10ex}|c|c|c}
     \toprule
     class name & edges & self-loops & multiple-edges & weighted\\
     \midrule
     \verb|SimpleGraph| & undirected & \xmark & \xmark & \xmark \\
     \verb|Multigraph| & undirected & \xmark & \cmark & \xmark \\
     \verb|Pseudograph| & undirected & \cmark & \cmark & \xmark\\
     \verb|DefaultUndirectedGraph| & undirected & \cmark & \xmark & \xmark\\
     \hline
     \verb|SimpleWeightedGraph| & undirected & \xmark & \xmark & \cmark\\
     \verb|WeightedMultigraph| & undirected & \xmark & \cmark & \cmark\\
     \verb|WeightedPseudograph| & undirected & \cmark & \cmark & \cmark\\
     \verb|DefaultUndirectedWeightedGraph| & undirected & \cmark & \xmark & \cmark\\   
     \hline     
     \verb|SparseIntUndirectedGraph| & undirected & \cmark & \cmark & \xmark\\     
     \verb|SparseIntUndirectedWeightedGraph| & undirected & \cmark & \cmark & \cmark\\     
     \hline
     \verb|SimpleDirectedGraph| & directed & \xmark & \xmark & \xmark\\
     \verb|DirectedMultigraph| & directed & \xmark & \cmark &\xmark\\
     \verb|DirectedPseudograph| & directed & \cmark & \cmark & \xmark\\
     \verb|DefaultDirectedGraph| & directed & \cmark & \xmark & \xmark\\   
     \hline
     \verb|SimpleDirectedWeightedGraph| & directed & \xmark & \xmark & \cmark\\
     \verb|DirectedWeightedMultigraph| & directed & \xmark & \cmark & \cmark\\
     \verb|DirectedWeightedPseudograph| & directed & \cmark & \cmark & \cmark\\
     \verb|DefaultDirectedWeightedGraph| & directed & \cmark & \xmark & \cmark\\
     \hline
     \verb|SparseIntDirectedGraph| & directed & \cmark & \cmark & \xmark\\     
     \verb|SparseIntDirectedWeightedGraph| & directed & \cmark & \cmark & \cmark\\     
     \hline     
     \bottomrule
  \end{tabular}
  \caption {All available graph implementation classes.}\label{table:graphclasses}
  \end{table}

\section{Algorithms} \label{sec:algorithms}

\jgrapht{} contains a large number of algorithms. A detailed discussion of each algorithm is outside the scope of
this paper; instead a general overview of the algorithms currently supported is provided.
Most of these algorithms are single-threaded, unless otherwise explicitly mentioned.

\paragraph{Connectivity}
Detecting connected components in graphs is a fundamental problem. For undirected graphs or weakly connected components in
directed graphs, standard traversals such as BFS or DFS suffice. For directed graphs, the library provides the linear time
algorithm of Kosaraju-Sharir~\cite{sharir1981strong} using 
two DFS traversals, as well as Gabow's algorithm~\cite{Gabow:2000:PDS:347744.347755}.
The classic Algorithm 447~\cite{hopcroft1973algorithm} is also provided for the computation of biconnected components. These
algorithms can also be used to identify \textit{cutpoints} and \textit{bridges} in a graph, or to construct a Block-Cutpoint graph.

\paragraph{LCA}
The least common ancestor of two nodes $v$ and $u$ in a tree or in a directed acyclic
graph (DAG) $T$, is the deepest node that has both $v$ and $u$ as descendants. A naive implementation (supporting both trees and DAGs) and the offine tree algorithm of Tarjan~\cite{gabow1983linear} can be used for small graph sizes or for batched queries. For larger tree instances, the library provides three additional implementations with different space time tradeoffs: (a) using the heavy-path decomposition with linear space to support LCA queries in $\Oh(\log n)$ time, (b) using the Euler-Tour technique~\cite{berkman1993recursive} and the classic reduction~\cite{bender2000lca} to the RMQ (range minimum query) problem to support LCA queries in $\Oh(1)$ time but with $\Oh(n \log n)$ space, and (c) a preprocessing approach which improves over the naive approach by computing jump pointers, using dynamic programming~\cite{bender2004level}. In the latter approach, each node stores jump pointers to ancestors at levels $1, 2, 4, \ldots, 2^k$.
Queries are answered by repeatedly jumping from node to node, each time jumping more than half of the remaining levels between the current ancestor and the goal ancestor (i.e. the lca). The worst-case number of jumps is $\Oh(\log n )$ which means that this method has $\Oh(\log n)$ query time again using $\Oh(n \log n)$ space.

\paragraph{Cycles}
Another fundamental problem involves enumerating all simple cycles of a graph. Several classic algorithms have been implemented
for this problem such as the algorithms of Tiernan~\cite{tiernan1970efficient}, Tarjan~\cite{tarjan1973enumeration},
Johnson~\cite{johnson1975finding}, Szwarcfiter and Lauer~\cite{szwarcfiter1976search}, and  Hawick and
James~\cite{hawick2008enumerating}. 

Additionally, the set of Eulerian subgraphs (subgraphs where all vertices have even degrees) forms the cycle space of a graph
(over the two-element finite field). A cycle basis is a basis of this vector space. The library contains a variant of Paton's
algorithm~\cite{paton1969algorithm} as well as some classic fundamental cycle basis construction algorithms using graph
traversals~\cite{deo1982algorithms}.
    
\paragraph{Shortest Paths}
The library contains extensive support for shortest path computations, both single-source and all-pairs. When all edge weights
are non-negative, Dijkstra's algorithm can be used. In \jgrapht{}, Dijkstra's algorithm is implemented using a Fibonacci
heap. A bidirectional variant is also included which enhances performance significantly for source-target queries. Additionally,
when edge weights can be negative, users can resort to the Bellman-Ford algorithm or Johnson's algorithm. Support for all-pairs
shortest paths is provided by the Floyd-Warshall algorithm.
The delta-stepping algorithm~\cite{MESA03}, a parallel algorithm for the single-source shortest path problem, is also included.

The library also contains an A* implementation together with the ALT admissible heuristic~\cite{goldberg2005computing} and
Martin's algorithm for the multi-objective shortest paths problem~\cite{martins1984multicriteria}.
With respect to $k$-shortest paths, \jgrapht{} includes 
variants of the Bellman-Ford algorithms for finding $k$-shortest simple paths, 
Eppstein's algorithm~\cite{eppstein1998finding} for finding the $k$ shortest paths between two vertices,
Yen's~\cite{martins2003new} algorithm for finding loop-less shortest paths and 
Suurballe and Tarjan~\cite{suurballe1984quick} algorithm for finding edge disjoint shortest paths.

Finally, a graph measurer class offers additional distance related metrics such as the graph diameter, the radius, vertex
eccentricities, the graph center, and graph (pseudo) periphery.

\paragraph{Node Centrality}
Node centrality measures the importance of nodes inside a network. Centrality metrics play a crucial role in social network analysis. The library support several vertex
centrality~\cite{newman2001scientific} metrics including alpha, betweenness, closeness, coreness, harmonic centrality and PageRank; 
see~\cite{newman2018networks} for details.
For betweenness centrality the algorithm of Brandes~\cite{brandes2001faster} is used. Coreness is computed using the techniques described in~\cite{matula1983smallest}. The remaining measures are computed using power iteration.

\paragraph{Spanning Trees and Spanners}
The minimum spanning tree problem asks to compute a spanning tree in a weighted graph of minimum total weight. The library includes Prim, Kruskal and Bor\r{u}vka's algorithms for the construction of minimum spanning trees. Prim's algorithm is implemented with a Fibonacci heap, while Kruskal's and Bor\r{u}vka's algorithms rely on a Union-Find data structure with union-by-rank and path-compression.
More general spanners can also be computed using for example the greedy algorithm for $(2k-1)$-multiplicative spanner construction~\cite{althofer1993sparse}.

\paragraph{Recognizing Graphs}
The library contains algorithms for the recognition of important types of graphs. Examples are bipartite graphs, chordal graphs
and Berge graphs. Bipartite graphs are recognized by standard graph traversals. For the recognition of chordal graphs we compute
a perfect elimination order either using {\em maximum cardinality search}~\cite{berry2004maximum} or {\em lexicographic breadth
first search}~\cite{corneil2004lexicographic}. Both require linear time. Finally, recognizing Berge graphs is accomplished
using the $\mathcal{O}(n^9)$ state-of-the-art algorithm of Chudnovsky et al.~\cite{chudnovsky2005recognizing}. Recall that a
graph is Berge if no induced subgraph of $G$ is an odd cycle of length at least five or the complement of such a cycle.

\paragraph{Matchings} Matching algorithms for general, bipartite, weighted and unweighted graphs are provided.
For maximum cardinality matching in general graphs the library includes the highly efficient $\Oh(m n \alpha(m,n))$
implementation of Edmonds~\cite{edmonds1965maximum} algorithm presented in the LEDA book~\cite{mehlhorn1995leda}.
For bipartite graphs, the user can invoke Hopcroft and Karp's algorithm~\cite{hopcroft1973n}.
To calculate a maximum weight matching in bipartite graphs, there is a highly efficient $\Oh(n(m+n \log n))$
implementation, again from the LEDA book. Minimum weight perfect bipartite matchings can be computed using
the $\Oh(n^3)$ Hungarian method. To compute a minimum weight perfect matching in general graphs, there is an efficient
implementation of Edmond's algorithm using the techniques introduced by the Blossom~V
implementation~\cite{kolmogorov2009blossom}. Finally, several fast $1/2$-approximation algorithms for matchings are provided,
including (a) a greedy algorithm and (b) the linear time path growing~\cite{drake2003simple} algorithm.

\paragraph{Cuts and Flows}
Maximum flows and minimum cuts in graphs are by definition closely related. The maximum flow problem~\cite{AMO93} involves calculating a feasible flow of maximum
value from a source vertex $s$ to a sink vertex $t$ through a capacitated network. Similarly, a minimum $s-t$ cut in a graph is a partitioning of the vertices $V$
into two disjoint subsets $S$ and $T$ such that $s\in S$, $t\in T$ while minimizing the sum of weights of the edges with exactly one endpoint in $S$ and one endpoint
in $T$. To efficiently calculate maximum $s-t$ flows, and by extension minimum $s-t$ cuts, the library provides implementations of the
Edmonds-Karp algorithm~\cite{edmonds1972theoretical}, the Push-Relabel algorithm~\cite{goldberg1988new}, and Dinic's algorithm~\cite{dinic1970algorithm}. 
Determining maximum $s-t$ flows or minimum $s-t$ cuts
for every $s-t$ pair in the graph can be realized by computing
resp.\ an Equivalent Flow tree~\cite{gusfield1990very} or a Gomory-Hu tree~\cite{gomory1961multi} using Gusfield's algorithm. The Gomory-Hu tree can also be used to
compute the minimum cut in the graph, i.e. the minimum cut over all $s-t$ pairs. Alternatively, the user can employ Stoer and Wagner's algorithm~\cite{stoer1997simple}
for this purpose.
Finally, the more general minimum-cost flow problem, which considers both costs and capacities for each arc in the network, can be solved by the successive shortest path algorithm, with or without capacity scaling~\cite{AMO93}. An implementation of the algorithm by Padberg and Rao~\cite{PARA82} to compute Odd Minimum Cut-Sets is also present.

\paragraph{Isomorphism}
(Sub)graph isomorphisms can be computed through the classic VF2~\cite{cordella2004sub} algorithm. 
Additionally, efficient heuristic isomorphic tests based on color
refinement~\cite{berkholz2017tight} are also provided.

\paragraph{Coloring}
The well-know NP-hard graph coloring problem entails the assignment of colors to vertices of a graph such that no two adjacent vertices share the same color. The library includes the exact coloring algorithm of Brown~\cite{brown1972chromatic} as well as several heuristic algorithms such as (a) greedy, (b) random greedy, (c) largest-degree-first (d) smallest-degree-last, and (e) saturation-degree~\cite{brelaz1979new} coloring.

\paragraph{Cliques}
The Bron-Kerbosch algorithm is an algorithm for enumerating all maximal cliques in an 
undirected graph. The library contains several variants. 
\begin{itemize} 
    \item Implementation of the Bron-Kerbosch clique enumeration algorithm as described in~\cite{samudrala1998graph}.
    \item Bron-Kerbosch maximal clique enumeration algorithm with pivot. The 
    pivoting follows the rule from Tomita et al.~\cite{tomita2006worst}, in which
    the authors show that this rule guarantees that the Bron-Kerbosch algorithm has worst-case running time $\Oh(3^{n/3})$, excluding time to write the output, where $n$ is the number of vertices of the graph; this is worst-case optimal.
    \item Bron-Kerbosch maximal clique enumeration algorithm with pivot and degeneracy ordering. The algorithm is a variant of the Bron-Kerbosch algorithm which apart from the pivoting uses a degeneracy ordering of the vertices. The algorithm is described in Eppstein et al.~\cite{eppstein2010listing} and has running time $\Oh(d n 3^{d/3})$ where $n$ is the number of vertices of the graph and $d$ is the degeneracy of the graph.
\end{itemize}
Moreover, algorithms to compute clique minimal separator decompositions~\cite{berry2010introduction} and maximum cliques in chordal graphs are also provided. 

\paragraph{Vertex Cover}
The minimum vertex cover problem is yet another classical NP-hard problem and involves selecting a subset of vertices of minimum cardinality such that each edge of the graph is incident to at least one selected vertex. \jgrapht{} provides
(a) an exact branch-and-bound algorithm, (b) a greedy heuristic and (c) various 2-approximation algorithms which differ either in running time or in solution quality, including the Bar-Yehuda and Even algorithm~\cite{bar1981linear} and Clarkson's algorithm~\cite{clarkson1983modification}.

\paragraph{Tours}
Several algorithms to compute tours, that is, both Hamiltonian Cycles (HCs) and Eulerian Cycles (ECs), are available. To solve the Traveling Salesman Problem (TSP) with optimality, thereby obtaining a minimum cost HC, the Held-Karp dynamic programming algorithm can be used. For weighted graphs satisfying the triangle inequality, two approximation algorithms are provided. The first algorithm is a 2-approximation and follows a traditional approach which first computes a MST, which is then traversed in a depth-first search manner to obtain a tour. The second algorithm is an implementation of Christofides $3/2$-approximation~\cite{CHRIS76}. Finally a 2-OPT heuristic is available to quickly compute HCs, but without any quality guarantees.
Determining whether a graph permits any HC irrespective of its cost remains an NP-Complete problem. Nevertheless, whenever the input graph satisfies Ore's condition, a HC can be identified in polynomial time ($\Oh(|V|^2)$) using Palmer's algorithm~\cite{palmer1997hidden}. Ore's condition essentially states that a graph with sufficiently many edges must contain a HC.

In addition to HCs, it is also possible to calculate ECs. ECs play an important role in the context of Arc Routing. To find
an EC in Eulerian graphs, Hierholzer's algorithm~\cite{hierholzer1873moglichkeit} can be used. Similarly, the Chinese Postman
Problem, requiring the calculation of a tour (closed walk) of minimum length which traverses every edge in a graph at least
once, can be solved efficiently using an implementation of Edmond's algorithm~\cite{EDMO73}. Obviously, when the input graph
is Eulerian, Edmond's algorithm returns an EC; otherwise the algorithm returns a closed walk of minimum length which traverses
some edges multiple times.

\section{Generators}\label{sec:generators}

\jgrapht{} provides a number of graph generators to deterministically generate graphs of arbitrary size which model and capture characteristics of real-world networks, e.g.\ social networks, communication networks, chemical interactions etc. These generators enable engineers and researchers to generate arbitrarily large synthetic datasets resembling real world data,
without the need to go through a costly and often time-consuming data collection process. Various types of graphs can be generated, including: complete
graphs, bipartite graphs, grid graphs, hypercubes, ring graphs, star graphs, wheel graphs, and others.
Additionally, dedicated generators for specific graphs famous in Graph Theory such as the Doyle graph,
the Petersen graph, Balaban graphs, etc are also provided.

Random graphs can be generated through the traditional $Gnm$ and $Gnp$ Erd{\"o}s-R{\'e}nyi~\cite{erdos1959random}
models. In the $Gnm$ model, a graph is chosen uniformly at random from the set of all graphs with $n$ nodes
and $m$ edges. In the $Gnp$ model a graph of $n$ nodes is constructed and each of the possible edges is chosen
with probability $p$. Similar models are available for the generation of random bipartite graphs where the
user specifies the size of the two partitions and either the number of edges or the edge probability.
Finally, generators for random regular graphs are also available.

More sophisticated models popular in e.g.\ social sciences are also provided.
The Barab{\'a}si-Albert~\cite{barabasi1999emergence} model starts from a small clique and incrementally
constructs a graph by adding new vertices one by one. Each new vertex is attached to a certain number
of previously constructed vertices using preferential attachment.
The Watts-Strogatz~\cite{albert2002statistical} model builds a graph by interpolating between a regular
lattice and a random graph. It starts from a regular lattice with $n$ nodes and $k \ll n$ edges per
node. Then it chooses a vertex and the edge that connects it to its nearest neighbor in a clockwise
sense. With probability $p$, it reconnects this edge to a vertex chosen uniformly at random over the
entire ring with duplicate edges forbidden; otherwise it leaves the edge in place. It continues this
process for each vertex of the ring and then repeats the procedure for the second-nearest neighbor, etc.
As there are $\frac{nk}{2}$ edges in the entire graph, the rewiring process
stops after $\frac{k}{2}$ laps. For intermediate values of $p$, the graph is a small-world network:
highly clustered like a regular graph, yet with small characteristic path length, like a random graph.
A small variant~\cite{newman1999renormalization} is also provided wherein instead of re-wiring, the
shortcut edges are added to the graph. This variant is sometimes called the Newman-Watts variant
of the Watts-Strogatz model.

The Kleinberg~\cite{kleinberg2000small} small-world model, which is also implemented, has as a basic
structure a two-dimensional grid and allows for edges to be directed. It begins with a set of nodes
(representing individuals in the social network) that are identified with the set of lattice points
in an $n \times n$ square. For a universal constant $p \geq 1$, the node $u$ has a directed edge
to every other node within lattice distance $p$ (these are its local contacts). For universal
constants $q \geq 0$ and $r \geq 0$, we also construct directed edges from $u$ to $q$ other nodes
(the long-range contacts) using independent random trials; the i-th directed edge from $u$ has
endpoint $v$ with probability proportional to $\frac{1}{d(u,v)^r}$
where $d(u,v)$ is the lattice distance from $u$ to $v$.

\section{Importing \& Exporting Graphs}\label{sec:io}

To increase interoperability between \jgrapht{} and other software solutions, and to facilitate
efficient storage of graphs, \jgrapht{} enables the user to read and write graphs in a variety of
popular data formats. Some of the common formats are: GML~\cite{himsolt1997gml}, CSV,
DIMACS~\cite{johnson1993network}, graph6 and sparse6~\cite{mckay2007graph6}. In particular, for
DIMACS the library supports the formats used in the 2nd challenge for max-clique problems and graph
coloring problems, as well as the shortest path format used in the 9th challenge. Sparse6 and graph6
are formats used for storing graphs in a compressed manner, using printable ASCII characters only.

Besides the aforementioned formats, \jgrapht{} also supports richer formats capable of storing
additional information such as graph attributes and labels. Among these formats are the DOT
language specification~\cite{gansner2000open} and GraphML~\cite{tamassia2013handbook}. Both
formats are fully supported. The implementations rely on Antlr v4~\cite{parr1995antlr} for
low-level parsing. For GraphML, two parsers are provided: one light-weight parser optimized
towards parsing speed, and one full fledged parser which implements the complete GraphML specifications.

\section{Experimental Evaluation}\label{sec:evaluation}

This section provides a computational evaluation of \jgrapht{}. In the evaluation, various algorithms
from \jgrapht{} (v1.4) are compared against their counterparts in alternative graph libraries. 
Given the large number of different algorithms and libraries, it is by no means possible to
provide an exhaustive comparison. Therefore, a number of commonly used algorithms and libraries have
been selected. 
These libraries were selected because of their popularity, plus the fact that they are open-source,
actively maintained and developed, and supporting a wide range of algorithms.

In particular, comparisons are made against \igraph{} (v0.7.1 written in C),
\bgl{} (v1.65 written in C++), \jung{} (v2.1.1 written in Java) and \nx{} (v2.1 written in Python).
The \igraph{} library represents graphs using a simple compressed-sparse-rows (CSR) based representation, which requires 
six different vectors, two of size $n$ and four with size $m$.
\bgl{}, on the other hand, implements graphs through adjacency lists and an adjacency matrix. \bgl{} enables the user
to customize the graph implementation through template parameters. In all our experiments we used the adjacency list graph parameterized 
with the STL vector container for both the vertex list and the edge lists containers. Moreover, in case of weighted graphs, edge weight properties are stored
directly on the edges as opposed to storing them in a separate table, thereby avoiding additional edge lookups.
Finally, the Java library \jung{} and the Python library \nx{}, similar to the default \jgrapht{}
implementation, rely on dictionaries to store the nodes and the node neighbors in a graph.
Most of the graph implementations in the \jung{} library are optimized for sparse graphs. 
In the experimental evaluation, we execute \jgrapht{} with the two different backends (Section~\ref{subsec:representations}):
the default backend, simply denoted
by `\jgrapht{}', and the sparse backend, denoted by `\jgraphtS{}'.

In addition to a computational study of different algorithmic implementations across libraries, we
conduct a limited internal comparison of different algorithms for the same fundamental mathematical
problem. Intrinsically, it is possible to compare algorithms by their worst-case runtime complexity,
but an experimental evaluation of their performance which largely depends on the quality
of their implementations is more informative in practice. This section is concluded by a comparison of
different graph representations, thereby evaluating speed and memory trade-offs between the different
representations.

\subsection{Instances}

For the computational evaluation, experiments are performed on a large number of benchmark instances. 
The instances are either taken from SNAP~\cite{leskovec2016snap}, or generated using the well-known 
(a) Barab{\'a}si-Albert model~\cite{barabasi1999emergence}, 
(b) Recursive Matrix (R-MAT) model~\cite{chakrabarti2004r}, and
(c) the $Gnp$ Erd{\"o}s-R{\'e}nyi~\cite{erdos1959random} random graph model.
Unless otherwise noted, we generated 10 different instances 
for each graph size and report results averaged over these 10 instances.
A table with the real-world instances from SNAP that were used in the experiments can be found in
Appendix~\ref{appendix:snap}.

Graphs following the Barab{\'a}si-Albert model are generated from a complete graph of size $m_0$.
New vertices are added to the graph, one by one, until a desired number of $n$ vertices is reached.
Each new node is connected to $m\leq m_0$ existing nodes with a probability that is proportional to
the number of links that the existing nodes already have. In the experiments, we set $m_0=20$, $m=10$,
for varying $n\in [0:250k]$. The resulting graphs are {\em sparse} and {\em scale-free}.\\
The recursive matrix (R-MAT) model generates a graph by 
recursively subdividing the adjacency matrix into four equal-sized partitions
and distributing edges within these four partitions using unequal probabilities
$a, b, c, d$ where $a+b+c+d=1$. Starting with an empty adjacency matrix, edges are 
added into the matrix one by one. The model generates very realistic graphs which have power-law
degree distributions and at the same time are small-world graphs.
We use the same settings as the Graph-500 benchmark\footnote{\url{https://graph500.org}}, i.e. for a 
given $SCALE$ the number of nodes equals $n=2^{SCALE}$, the number of edges $m=16 n$ and we set
$a=0.57, b=0.19, c=0.19$ and $d=0.05$.\\
In $Gnp$ Erd{\"o}s-R{\'e}nyi graphs, edges are included with probability $p$. Therefore, $Gnp$ graphs with
$n$ vertices have an expected number of edges equal to $p {n \choose 2}$. In the experiments, we use
$p=0.1$, thereby obtaining relatively {\em dense} graphs.

\subsection{Setup \& Measuring Methodology}

All experiments were executed on an Intel(R) Core(TM) i7-6700 CPU @ 3.4GHz running a 64-bit version of
GNU/Linux using 32GB of memory. To facilitate a fair comparison, all experiments were performed on a single
processor core. The algorithms written in \igraph{} and \bgl{} were compiled using GCC v7.3 using
the \verb|-O3| optimization flag. The Java libraries were executed using Java version 1.8.0\_221
on Java HotSpot(TM) 64-Bit GraalVM EE 19.2.0.1 (build 25.221-b11-jvmci-19.2-b02, mixed mode). We use
GraalVM~\cite{wurthinger2013one} in order
to compile all Java code ahead-of-time (AOT) into native executables, resulting in faster 
startup times and much lower runtime memory overheads. 
GraalVM's tool {\em native-image} produces a native executable with a lightweight subsystem called
SubstrateVM which includes all the necessary components like memory management,
thread scheduling, garbage collector, etc., in order to allow native execution without the use of the Java VM.
In order to avoid any unnecessary interference from garbage collection,
we used \verb|-Xmx30g| when executing any Java code. Python 3.6.5 was used for the execution of the \nx{}
library. All reported times are {\em wall-clock} times.

The use of ahead-of-time compilation makes it considerably easier to compare different libraries. Nevertheless, we still
have libraries written in C/C++ which explicitly cleanup memory and libraries written in Java and Python which rely on a 
garbage collector. In languages which include a garbage collector, we use the following procedure to measure the
running time of an algorithm:
(a) The graph is first imported from the input file and read into memory using the corresponding library, 
(b) at this point we explicitly call the garbage collector, 
(c) the start time is recorded, 
(d) the algorithm is then executed,
(e) we again explicitly call the garbage collector without releasing any references to the graph, and
(f) we record the finish time.
Our goal is to solely measure the execution time of the algorithm plus the time it takes to allocate and cleanup any
auxiliary data structures used by the algorithm.
In case of languages like C++ where memory is released when a "smart" pointer gets out of scope, we enclose the algorithm's 
execution inside an additional block statement, thus forcing any auxiliary memory to be released at the end of the block, 
just before we measure the finish time.

\subsection{Computational Results - External comparison}

In this subsection, algorithms from \jgrapht{} are compared against implementations from alternative
libraries. In particular the comparison uses the following algorithms: Dijkstra shortest path, PageRank,
Maximum Cardinality and Minimum Weighted Perfect Matching. 

%
%
\pgfplotscreateplotcyclelist{mybw}{%
dashdotdotted, every mark/.append style={solid},mark=star\\%
loosely dotted, every mark/.append style={solid, fill=gray}, mark=triangle*\\%
densely dashdotted,every mark/.append style={solid, fill=gray},mark=diamond*\\%
solid, every mark/.append style={solid, fill=gray}, mark=*\\%
dotted, every mark/.append style={solid, fill=gray}, mark=square*\\%
densely dotted, every mark/.append style={solid, fill=gray}, mark=otimes*\\%
dashed, every mark/.append style={solid, fill=gray},mark=diamond*\\%
loosely dashed, every mark/.append style={solid, fill=gray},mark=*\\%
densely dashed, every mark/.append style={solid, fill=gray},mark=square*\\%
dashdotted, every mark/.append style={solid, fill=gray},mark=otimes*\\%
}

%
%
\begin{figure}[t!]
  \begin{tikzpicture}
    
    \begin{groupplot}[
          group style={
            group size=2 by 2,
            vertical sep=1.5cm,
          },          
          height=0.25\textheight,
          width=0.50\textwidth, 
          grid=major, 
          grid style={dashed,gray!30}, 
          xlabel=nodes,
          xticklabel style={rotate=45,anchor=east,font=\scriptsize},
          yticklabel style={font=\scriptsize},
          scaled x ticks=false,
          cycle list name=mybw,
          ymode=log,
          legend columns=6,
          legend style={
            at={(1.0,2.5)},
            anchor=south east,
            /tikz/column 2/.style={column sep=5pt},
          }
      ]
      \nextgroupplot[
        ylabel=time,
        y unit=ms,
      ]
      \addplot
      table[x=nodes,y=time (ms),col sep=semicolon] {plots.r1/barabasi-albert-20-10-jgrapht-dijkstra.csv}; 

      \addplot
      table[x=nodes,y=time (ms),col sep=semicolon] {plots.r1/barabasi-albert-20-10-jgrapht-sparse-dijkstra.csv}; 

      \addplot
      table[x=nodes,y=time (ms),col sep=semicolon] {plots.r1/barabasi-albert-20-10-jung-dijkstra.csv}; 

      \addplot
      table[x=nodes,y=time (ms),col sep=semicolon] {plots.r1/barabasi-albert-20-10-networkx-dijkstra.csv};
          
      \addplot
      table[x=nodes,y=time (ms),col sep=semicolon] {plots.r1/barabasi-albert-20-10-BGL-dijkstra.csv}; 

      \addplot
      table[x=nodes,y=time (ms),col sep=semicolon] {plots.r1/barabasi-albert-20-10-igraph-dijkstra.csv}; 

      \nextgroupplot
      \addplot
      table[x=nodes,y=time (ms),col sep=semicolon] {plots.r1/gnp-0.1-jgrapht-dijkstra.csv};

      \addplot
      table[x=nodes,y=time (ms),col sep=semicolon] {plots.r1/gnp-0.1-jgrapht-sparse-dijkstra.csv};

      \addplot
      table[x=nodes,y=time (ms),col sep=semicolon] {plots.r1/gnp-0.1-jung-dijkstra.csv};

      \addplot
      table[x=nodes,y=time (ms),col sep=semicolon] {plots.r1/gnp-0.1-networkx-dijkstra.csv};

      \addplot 
      table[x=nodes,y=time (ms),col sep=semicolon] {plots.r1/gnp-0.1-BGL-dijkstra.csv};

      \addplot
      table[x=nodes,y=time (ms),col sep=semicolon] {plots.r1/gnp-0.1-igraph-dijkstra.csv};

      \nextgroupplot[
        ylabel=time,
        y unit=ms,
        xmode=log,
      ]

      \addplot
      table[x=nodes,y=time (ms),col sep=semicolon] {plots.r1/rmat-undirected-jgrapht-dijkstra.csv};

      \addplot
      table[x=nodes,y=time (ms),col sep=semicolon] {plots.r1/rmat-undirected-jgrapht-sparse-dijkstra.csv};

      \addplot
      table[x=nodes,y=time (ms),col sep=semicolon] {plots.r1/rmat-undirected-jung-dijkstra.csv};

      \addplot
      table[x=nodes,y=time (ms),col sep=semicolon] {plots.r1/rmat-undirected-networkx-dijkstra.csv};

      \addplot
      table[x=nodes,y=time (ms),col sep=semicolon] {plots.r1/rmat-undirected-BGL-dijkstra.csv};

      \addplot
      table[x=nodes,y=time (ms),col sep=semicolon] {plots.r1/rmat-undirected-igraph-dijkstra.csv};

      \nextgroupplot

      \addplot
      table[x=nodes,y=time (ms),col sep=semicolon] {plots.r1/USA-jgrapht-dijkstra.csv};

      \addplot
      table[x=nodes,y=time (ms),col sep=semicolon] {plots.r1/USA-jgrapht-sparse-dijkstra.csv};

      \addplot
      table[x=nodes,y=time (ms),col sep=semicolon] {plots.r1/USA-jung-dijkstra.csv};

      \addplot
      table[x=nodes,y=time (ms),col sep=semicolon] {plots.r1/USA-networkx-dijkstra.csv};

      \addplot
      table[x=nodes,y=time (ms),col sep=semicolon] {plots.r1/USA-BGL-dijkstra.csv};

      \addplot
      table[x=nodes,y=time (ms),col sep=semicolon] {plots.r1/USA-igraph-dijkstra.csv};

      \legend{\jgrapht{}, \jgraphtS{}, \jung{}, \nx{}, \bgl{}, \igraph{}}          
    \end{groupplot}
  \end{tikzpicture}
  \caption{Execution time of Dijkstra algorithm implementation in the
  five libraries using Barabasi-Albert (top-left), 
  $Gnp$ with $p=0.1$ (top-right), 
  undirected R-MAT ($a=0.57, b=0.19, c=0.19$) (bottom-left), 
  and USA roadmaps (bottom-right).
  }\label{elapsed:dijkstra}
\end{figure}
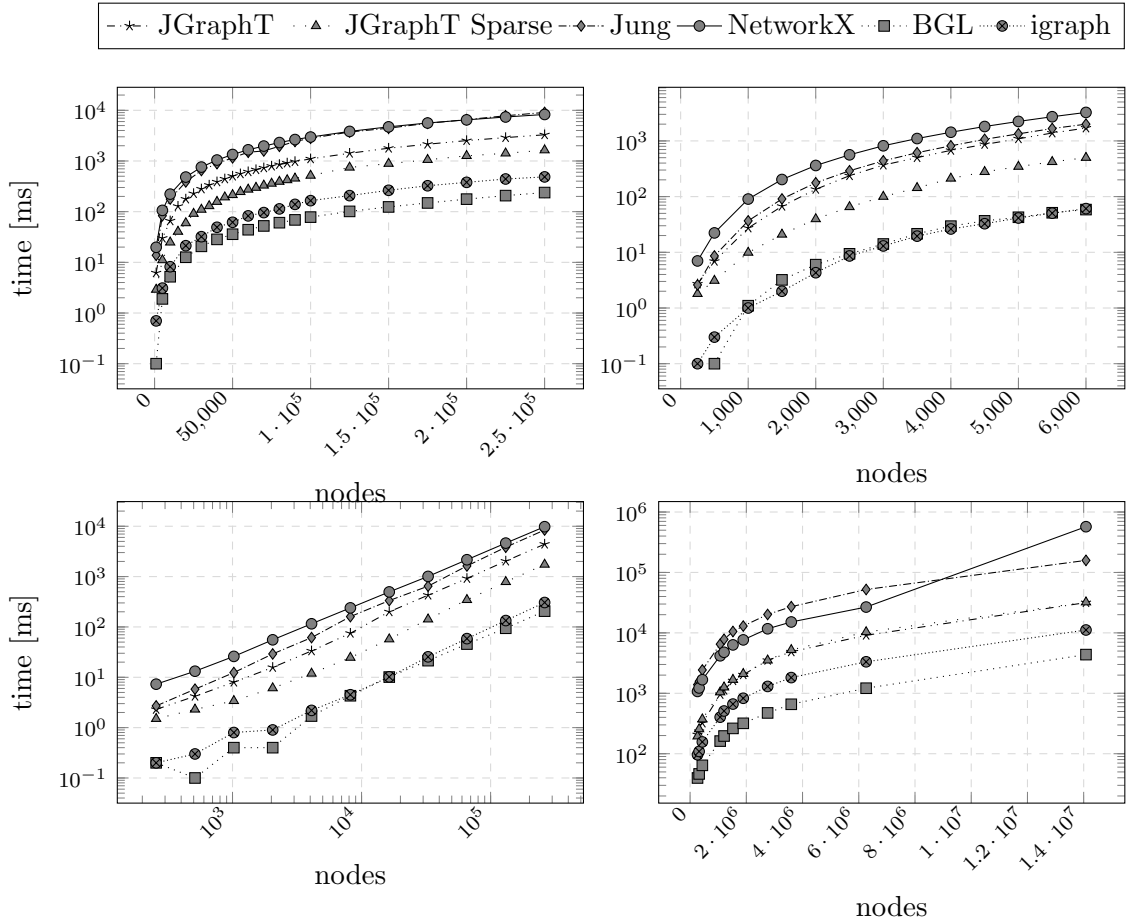

\subsubsection{Dijkstra Shortest Path} \label{subsub:dijkstra}

Dijkstra Shortest Path algorithm computes shortest paths from a single source node to all other nodes in the graph,
thereby producing a shortest-path tree. 
Figure~\ref{elapsed:dijkstra} compares the performance of the Dijkstra's shortest path implementations for \jgrapht{},
\jung{}, \nx{}, \bgl{} and \igraph{}. 
We used \jgrapht{}'s Dijkstra implementation using a $4$-ary heap. \bgl{} does the same while the remaining libraries 
use a binary heap.

The experiments are performed in both generated graphs and real-world USA 
road networks taken from the 9th DIMACS challenge~\cite{demetrescu2009shortest}.
For the generated instances, we executed Dijkstra's algorithm by starting from the same
node in each graph and computed the shortest path tree to all other vertices in the graph. The result is the average
running time over 10 different graphs constructed using the same parameters. 
For the road networks we picked uniformly at random ten source vertices and constructed one shortest path tree for
each source vertex. The result is the average running time over these 10 executions.

As can be observed from Figure~\ref{elapsed:dijkstra}, the C/C++ libraries \bgl{} and \igraph{} provide the best
performance. On average, measured over all four graphs, these libraries are resp.~7.3 and 3.0 times faster than
\jgraphtS{}. In all cases, the Java library \jung{} and the Python library \nx{} provide the worst performance
(resp.~5.2 and 10.5 times slower than \jgraphtS{}). In general, the \jgraphtS{} backend outperforms the more
flexible default backend.

In order to better understand and explain the performance difference between the C/C++ libraries and \jgraphtS{} we 
performed the following experiment. We used the \igraph{} library as a baseline and modified our implementation by 
gradually eliminating differences. As a first step we wrote a \jgrapht{} backend which uses exactly the same 
low level representation that \igraph{} is using. Afterwards, like \igraph{} does, we modified our Dijkstra
implementation to utilise the fact that vertices are integers and store distances and predecessor information directly
in arrays. Finally, we used a $d$-ary heap for the priority queue just like \igraph{} does. 
After all these modifications we rerun our experiments comparing our new implementation with \igraph{}. The performance 
difference in these experiments was the same as in Figure~\ref{elapsed:dijkstra}. Given the fact that the two
implementations of Dijkstra are almost identical, they both utilize the same graph 
representation, both use arrays and random access for distances and predecessor pointers, and both use similar heap
implementations, we conclude that the performance difference between \igraph{} and \jgraphtS{} is mostly due to the
extra overhead that GraalVM entails. 
Recall that GraalVM is a relatively new implementation, which is likely not able to produce the same quality 
executables as GCC at least w.r.t.\ the level of code optimization. Additionally, the SubstrateVM, while much more 
lightweight in comparison to a full-blown Java VM, still needs to perform additional bookeeping and thus entails 
additional performance penalties. This conclusion is further supported by our next experiment.

%
%
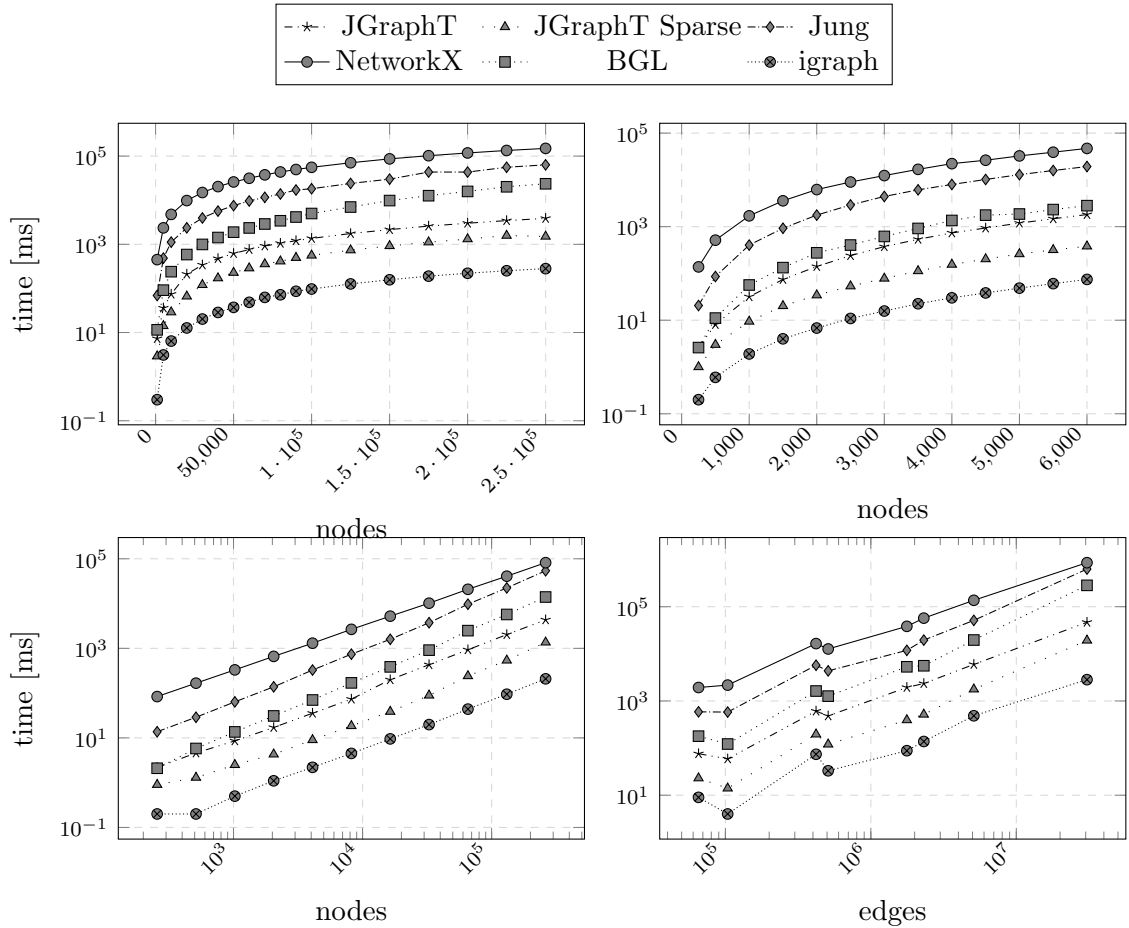
\begin{figure}[t!]
  \begin{tikzpicture}
    
    \begin{groupplot}[
          group style={
            group size=2 by 2,
            vertical sep=1.5cm,
          },
          height=0.25\textheight,
          width=0.50\textwidth, 
          grid=major, 
          grid style={dashed,gray!30}, 
          xlabel=nodes,
          xticklabel style={rotate=45,anchor=east,font=\scriptsize},
          yticklabel style={font=\scriptsize},
          scaled x ticks=false,
          cycle list name=mybw,
          ymode=log,
          legend columns=3,
          legend style={
            at={(0.5,2.5)},
            anchor=south east,
            /tikz/column 2/.style={column sep=5pt},
          }
      ]
      \nextgroupplot[
        ylabel=time,
        y unit=ms,
      ]
      \addplot
      table[x=nodes,y=time (ms),col sep=semicolon] {plots.r1/barabasi-albert-20-10-jgrapht-pagerank.csv};

      \addplot
      table[x=nodes,y=time (ms),col sep=semicolon] {plots.r1/barabasi-albert-20-10-jgrapht-sparse-pagerank.csv}; 

      \addplot
      table[x=nodes,y=time (ms),col sep=semicolon] {plots.r1/barabasi-albert-20-10-jung-pagerank.csv}; 

      \addplot
      table[x=nodes,y=time (ms),col sep=semicolon] {plots.r1/barabasi-albert-20-10-networkx-pagerank.csv};

      \addplot
      table[x=nodes,y=time (ms),col sep=semicolon] {plots.r1/barabasi-albert-20-10-BGL-pagerank.csv}; 

      \addplot
      table[x=nodes,y=time (ms),col sep=semicolon] {plots.r1/barabasi-albert-20-10-igraph-pagerank.csv};

      \nextgroupplot
          \addplot
          table[x=nodes,y=time (ms),col sep=semicolon] {plots.r1/gnp-0.1-jgrapht-pagerank.csv};

          \addplot
          table[x=nodes,y=time (ms),col sep=semicolon] {plots.r1/gnp-0.1-jgrapht-sparse-pagerank.csv};

          \addplot
          table[x=nodes,y=time (ms),col sep=semicolon] {plots.r1/gnp-0.1-jung-pagerank.csv};

          \addplot
          table[x=nodes,y=time (ms),col sep=semicolon] {plots.r1/gnp-0.1-networkx-pagerank.csv};

          \addplot 
          table[x=nodes,y=time (ms),col sep=semicolon] {plots.r1/gnp-0.1-BGL-pagerank.csv};

          \addplot
          table[x=nodes,y=time (ms),col sep=semicolon] {plots.r1/gnp-0.1-igraph-pagerank.csv};

      \nextgroupplot[
        ylabel=time,
        y unit=ms,
        xmode=log,
      ]
          \addplot
          table[x=nodes,y=time (ms),col sep=semicolon] {plots.r1/rmat-jgrapht-pagerank.csv};

          \addplot
          table[x=nodes,y=time (ms),col sep=semicolon] {plots.r1/rmat-jgrapht-sparse-pagerank.csv};

          \addplot
          table[x=nodes,y=time (ms),col sep=semicolon] {plots.r1/rmat-jung-pagerank.csv};

          \addplot
          table[x=nodes,y=time (ms),col sep=semicolon] {plots.r1/rmat-networkx-pagerank.csv};

          \addplot 
          table[x=nodes,y=time (ms),col sep=semicolon] {plots.r1/rmat-BGL-pagerank.csv};

          \addplot
          table[x=nodes,y=time (ms),col sep=semicolon] {plots.r1/rmat-igraph-pagerank.csv};

          \nextgroupplot[
            xlabel=edges,
            xmode=log,
          ]
          \addplot
          table[x=nodes,y=time (ms),col sep=semicolon] {plots.r1/snap-directed-jgrapht-pagerank.csv};

          \addplot
          table[x=nodes,y=time (ms),col sep=semicolon] {plots.r1/snap-directed-jgrapht-sparse-pagerank.csv};

          \addplot
          table[x=nodes,y=time (ms),col sep=semicolon] {plots.r1/snap-directed-jung-pagerank.csv};

          \addplot
          table[x=nodes,y=time (ms),col sep=semicolon] {plots.r1/snap-directed-networkx-pagerank.csv};

          \addplot 
          table[x=nodes,y=time (ms),col sep=semicolon] {plots.r1/snap-directed-BGL-pagerank.csv};

          \addplot
          table[x=nodes,y=time (ms),col sep=semicolon] {plots.r1/snap-directed-igraph-pagerank.csv};          
          
          \legend{\jgrapht{}, \jgraphtS{}, \jung{}, \nx{}, \bgl{}, \igraph{}}          
    \end{groupplot}
  \end{tikzpicture}
  \caption{Execution time of PageRank algorithm implementation in the
  five libraries using Barabasi-Albert (up-left), 
   $Gnp$ with $p=0.1$ (up-right), 
   directed R-MAT ($a=0.57, b=0.19, c=0.19$) (bottom-left), 
   and SNAP directed graphs (bottom-right).
  }\label{elapsed:pagerank}
\end{figure}

\subsubsection{PageRank} \label{subsub:pagerank}
Pagerank is regularly used in bibliometrics, social and information network analysis, and for link
prediction and recommendation~\cite{GLEI15}. For all libraries, we execute the PageRank algorithm with
a damping factor of $0.85$, $20$ iterations and tolerance equal to $10^{-16}$. While some libraries,
such as \igraph{}, contain several alternative PageRank implementations, we selected the implementation
based on power iterations, as the same technique is used by the other libraries.

Similar to the computational results of Dijkstra's Shortest Path algorithm, the best performance is again
obtained with \igraph{} (6.0 times faster than \jgrapht{} Sparse), but \bgl{}, \jung{} and \nx{} are
resp.\ 13.4, 34, 62.4 times slower than \jgrapht{} Sparse. These performance differences are consistent among
the 4 graph types. Interesting to observe is that also \jgrapht{} with its default backend which relies on
hashtables to perform edge and vertex lookups, outperforms the vector based \bgl{} implementation. A detailed
code comparison of the Pagerank implementations reveals that the lower performance of \bgl{} can be attributed
to the fact that \bgl{} repeatedly reads the graph while executing the iterations of the Pagerank algorithm,
whereas both \igraph{} and \jgrapht{} read the graph only once by transforming it into a more appropriate integer 
based matrix representation and then run PageRank directly on top of this matrix.

These results further support our claim that compared with the C/C++ libraries, when the algorithmic details are
similar, the performance difference is mostly due to the different programming language implementations.

\subsubsection{Maximum Cardinality \& Minimum Weight Perfect Matchings}

Graph matching is a fundamental problem in computer science and graph theory, and has applications in computer vision, computational biology,
arc routing and pattern recognition. In this subsection we evaluate the performance of algorithms for the Maximum Cardinality Matching Problem
(MCMP) and the Minimum Weight Perfect Matching Problem (MWPMP) in general graphs. While both problems admit very efficient algorithms, their
implementations are highly complex and require a significant amount of engineering. Consequently, there exist only a few commercial and
non-commercial libraries which incorporate implementations for either of these problems.

Matching problems can be straightforwardly formulated as Integer Linear Programming Problems (ILPs), which can be solved by any off-the-shelve
ILP solver. Given an undirected graph $G(V,E)$ with vertex set $V$, edge set $E\subseteq V\times V$, and edge weights $c_{ij}$ for all $(i,j)\in E$,
the MCMP and MWPMP can be modeled as ILPs through Equations~\eqref{eq:matching.0}-\eqref{eq:matching.2}
and~\eqref{eq:matching.3}-\eqref{eq:matching.5} as follows:

\begin{minipage}{.45\linewidth}
\begin{flalign}
\mbox{MCMP:}&&\nonumber\\
\label{eq:matching.0} \mbox{max. } & \sum_{(i,j)\in E}x_{ij}&\\
\label{eq:matching.1} \mbox{s.t. } &\sum_{j : (i,j)\in E}x_{ij}\leq 1 &\forall i \in V\\ 
\label{eq:matching.2} &x_{ij}\in \{0,1\} &\forall (i,j)\in E
\end{flalign}
\end{minipage}%
\hfill
\begin{minipage}{.45\linewidth}
\begin{flalign}
\mbox{MWPMP:}&&\nonumber\\
\label{eq:matching.3} \mbox{min. } & \sum_{(i,j)\in E}c_{ij} x_{ij}&\\
\label{eq:matching.4} \mbox{s.t. } &\sum_{j : (i,j)\in E}x_{ij}=1 &\forall i \in V\\ 
\label{eq:matching.5} &x_{ij}\in \{0,1\} &\forall (i,j)\in E
\end{flalign}
\end{minipage}

\medskip

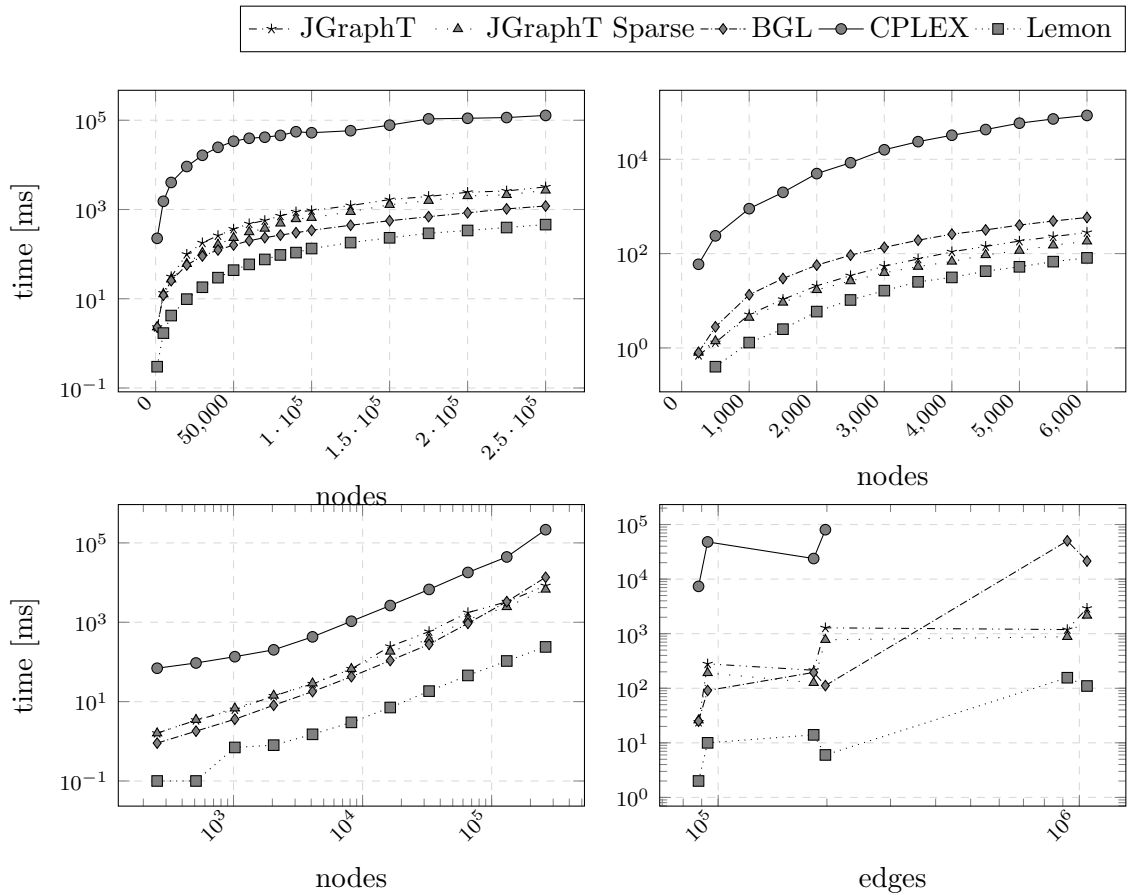
\begin{figure}[t!]
  \begin{tikzpicture}
    
    \begin{groupplot}[
          group style={
            group size=2 by 2,
            vertical sep=1.5cm,
          },
          height=0.25\textheight,
          width=0.50\textwidth, 
          grid=major, 
          grid style={dashed,gray!30}, 
          xlabel=nodes,
          xticklabel style={rotate=45,anchor=east,font=\scriptsize},
          yticklabel style={font=\scriptsize},
          scaled x ticks=false,
          cycle list name=mybw,
          ymode=log,
          legend columns=5,
          legend style={
            at={(1.0,2.5)},
            anchor=south east,
            /tikz/column 2/.style={column sep=5pt},
          }
      ]
      \nextgroupplot[
        ylabel=time,
        y unit=ms,
      ]

      \addplot
      table[x=nodes,y=time (ms),col sep=semicolon] {plots.r1/barabasi-albert-20-10-jgrapht-mcm.csv};

      \addplot
      table[x=nodes,y=time (ms),col sep=semicolon] {plots.r1/barabasi-albert-20-10-jgrapht-sparse-mcm.csv};

      \addplot
      table[x=nodes,y=time (ms),col sep=semicolon] {plots.r1/barabasi-albert-20-10-BGL-mcm.csv};

      \addplot
      table[x=nodes,y=time (ms),col sep=semicolon] {plots.r1/barabasi-albert-20-10-cplex-mcm.csv};

      \addplot 
      table[x=nodes,y=time (ms),col sep=semicolon] {plots.r1/barabasi-albert-20-10-lemon-mcm.csv};      

      \nextgroupplot
      \addplot
      table[x=nodes,y=time (ms),col sep=semicolon] {plots.r1/gnp-0.1-jgrapht-mcm.csv}; 

      \addplot
      table[x=nodes,y=time (ms),col sep=semicolon] {plots.r1/gnp-0.1-jgrapht-sparse-mcm.csv}; 

      \addplot
      table[x=nodes,y=time (ms),col sep=semicolon] {plots.r1/gnp-0.1-BGL-mcm.csv}; 

      \addplot
      table[x=nodes,y=time (ms),col sep=semicolon] {plots.r1/gnp-0.1-cplex-mcm.csv};
          
      \addplot
      table[x=nodes,y=time (ms),col sep=semicolon] {plots.r1/gnp-0.1-lemon-mcm.csv}; 

      \nextgroupplot[
        ylabel=time,
        y unit=ms,
        xmode=log,
      ]

      \addplot
      table[x=nodes,y=time (ms),col sep=semicolon] {plots.r1/rmat-undirected-jgrapht-mcm.csv};

      \addplot
      table[x=nodes,y=time (ms),col sep=semicolon] {plots.r1/rmat-undirected-jgrapht-sparse-mcm.csv};

      \addplot
      table[x=nodes,y=time (ms),col sep=semicolon] {plots.r1/rmat-undirected-BGL-mcm.csv};

      \addplot
      table[x=nodes,y=time (ms),col sep=semicolon] {plots.r1/rmat-undirected-cplex-mcm.csv};

      \addplot 
      table[x=nodes,y=time (ms),col sep=semicolon] {plots.r1/rmat-undirected-lemon-mcm.csv};      

      \nextgroupplot[
        xlabel=edges,
        xmode=log,
      ]
      \addplot
      table[x=nodes,y=time (ms),col sep=semicolon] {plots.r1/snap-undirected-jgrapht-mcm.csv}; 

      \addplot
      table[x=nodes,y=time (ms),col sep=semicolon] {plots.r1/snap-undirected-jgrapht-sparse-mcm.csv};

      \addplot
      table[x=nodes,y=time (ms),col sep=semicolon] {plots.r1/snap-undirected-BGL-mcm.csv}; 

      \addplot
      table[x=nodes,y=time (ms),col sep=semicolon] {plots.r1/snap-undirected-cplex-mcm.csv};
          
      \addplot
      table[x=nodes,y=time (ms),col sep=semicolon] {plots.r1/snap-undirected-lemon-mcm.csv}; 
      \legend{\jgrapht{}, \jgraphtS{}, \bgl{}, \cplex{}, \lemon{}} 
    \end{groupplot}
  \end{tikzpicture}
  \caption{Execution time of the maximum cardinality matching algorithm implementation in
  different libraries using Barabasi-Albert (up-left), 
   $Gnp$ with $p=0.1$ (up-right), 
   undirected R-MAT ($a=0.57, b=0.19, c=0.19$) (bottom-left), 
   and SNAP undirected graphs (bottom-right).
  }\label{elapsed:mcm}
\end{figure}

To provide a point of reference, as part of our computational study, we solve these models with the commercial ILP solver
ILOG CPLEX 12.8, and compare against a number of dedicated matching algorithms. In these experiments, CPLEX is invoked with
default parameters. 
Figure~\ref{elapsed:mcm} compares the execution of the MCMP implementations of \jgrapht{}, \bgl{}, and \lemon{}
(v1.3.1 written in C++). \igraph{}, \jung{} and \nx{} have been omitted from the comparisons as neither includes an MCMP
implementation.
Similarly, CPLEX results for the largest graphs have been omitted since the solver ran out of memory.

As can be observed from Figure~\ref{elapsed:mcm} the dedicated MCMP implementations are significantly faster than the
generic ILP solver CPLEX on all graphs: CPLEX is about 57.4 times slower than \jgrapht{}, with the differences being
bigger for denser graphs. The \bgl{} is slightly faster than \jgrapht{} on Barabasi-Albert graphs, but performs worse
on the real-world graphs from SNAP and dense $Gnp$ graphs.
Averaged over all graphs, \jgrapht{} is 5.7 times faster than \bgl{}. The best results are however obtained with the
C++ library lemon which is on average 9.7 times faster than \jgrapht{}.

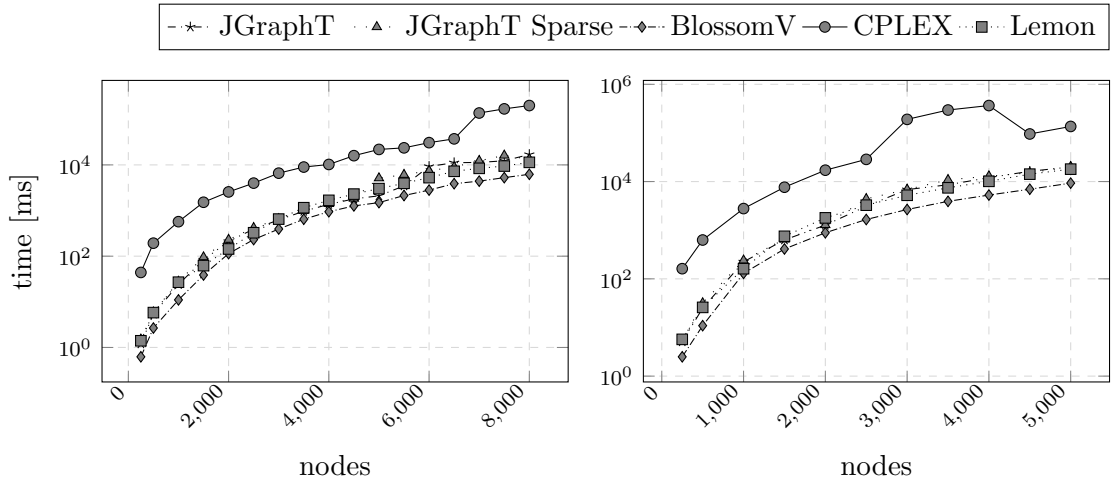
\begin{figure}[t!]
  \begin{tikzpicture}
    
    \begin{groupplot}[
          group style={group size=2 by 1},
          height=0.25\textheight,
          width=0.5\textwidth, 
          grid=major, 
          grid style={dashed,gray!30}, 
          xlabel=nodes,
          xticklabel style={rotate=45,anchor=east,font=\scriptsize},
          yticklabel style={font=\scriptsize},
          scaled x ticks=false,
          cycle list name=mybw,
          ymode=log,
          legend columns=6,
          legend style={
            at={(1.0,1.1)},
            anchor=south east,
            /tikz/column 2/.style={column sep=5pt},
          }
      ]
      \nextgroupplot[
        ylabel=time,
        y unit=ms,
      ]

      \addplot
      table[x=nodes,y=time (ms),col sep=semicolon] {plots.r1/gnp-0.1-jgrapht-mwpm.csv};

      \addplot
      table[x=nodes,y=time (ms),col sep=semicolon] {plots.r1/gnp-0.1-jgrapht-sparse-mwpm.csv};

      \addplot
      table[x=nodes,y=time (ms),col sep=semicolon] {plots.r1/gnp-0.1-blossom5-mwpm.csv};

      \addplot
      table[x=nodes,y=time (ms),col sep=semicolon] {plots.r1/gnp-0.1-cplex-mwpm.csv};

      \addplot 
      table[x=nodes,y=time (ms),col sep=semicolon] {plots.r1/gnp-0.1-lemon-mwpm.csv};

      \nextgroupplot
      \addplot
      table[x=nodes,y=time (ms),col sep=semicolon] {plots.r1/gnp-0.5-jgrapht-mwpm.csv}; 

      \addplot
      table[x=nodes,y=time (ms),col sep=semicolon] {plots.r1/gnp-0.5-jgrapht-sparse-mwpm.csv}; 

      \addplot
      table[x=nodes,y=time (ms),col sep=semicolon] {plots.r1/gnp-0.5-blossom5-mwpm.csv}; 

      \addplot
      table[x=nodes,y=time (ms),col sep=semicolon] {plots.r1/gnp-0.5-cplex-mwpm.csv};
          
      \addplot
      table[x=nodes,y=time (ms),col sep=semicolon] {plots.r1/gnp-0.5-lemon-mwpm.csv}; 

      \legend{\jgrapht{}, \jgraphtS{}, \blossom{}, \cplex{}, \lemon{}}

    \end{groupplot}
  \end{tikzpicture}
  \caption{Minimum Weight Perfect Matching. Left is $Gnp$ with $p=0.1$, right is $Gnp$ with $p=0.5$. 
  }\label{elapsed:mwpm}
\end{figure}

Figure~\ref{elapsed:mwpm} contains a comparison of the most efficient algorithmic implementations available for the MWPMP.
In order to generate
random instances which are guaranteed to contain a perfect matching, similar to the $Gnp$ model, we first created $n$ vertices,
connected them in pairs with edges and then created all remaining edges with probability equal to $p$. Notice that the \bgl{}
library does not provide a MWPMP implementation and is therefore excluded from the comparison. Instead we included the 
\blossom{}\footnote{\url{http://pub.ist.ac.at/~vnk/software/blossom5-v2.05.src.tar.gz}}~\cite{kolmogorov2009blossom}
implementation (v2.05 written in \cpp{}) which is currently considered the fastest MWPMP solver available. As can be observed
from Figure~\ref{elapsed:mwpm}, \jgrapht{} is highly competitive, even when compared with the state-of-the-art low
level \blossom{} implementation. \blossom{} and Lemon are resp. 2.4 and 1.3 times faster than \jgrapht{}. All methods are more
than an order of magnitude faster than CPLEX.
Finally, note that for this particular algorithm, there is no significant difference in performance between the default and the
sparse backend of \jgrapht{}. The reason for this is that the algorithm reads the graph only once, and maintains its own
internal representation. 

\begin{figure}[t!]
  \begin{tikzpicture}
    
    \begin{groupplot}[
          group style={
            group size=2 by 2,
            vertical sep=1.5cm,
          },          
          height=0.25\textheight,
          width=0.50\textwidth, 
          grid=major, 
          grid style={dashed,gray!30}, 
          xlabel=nodes,
          xticklabel style={rotate=45,anchor=east,font=\scriptsize},
          yticklabel style={font=\scriptsize},
          scaled x ticks=false,
          cycle list name=mybw,          
          ymode=log,
          legend columns=1,
          legend style={
            font=\scriptsize,
            at={(0.95,0.05)},
            anchor=south east,
            /tikz/column 2/.style={column sep=1pt},
          }
      ]
      \nextgroupplot[
        ylabel=time,
        y unit=ms,
      ]
      \addplot 
      table[x=nodes,y=time (ms),col sep=semicolon] {plots.r1/barabasi-albert-20-10-jgrapht-sparse-prim.csv};
      \addplot
      table[x=nodes,y=time (ms),col sep=semicolon] {plots.r1/barabasi-albert-20-10-jgrapht-sparse-kruskal.csv};
      \addplot
      table[x=nodes,y=time (ms),col sep=semicolon] {plots.r1/barabasi-albert-20-10-jgrapht-sparse-boruvka.csv};

      \nextgroupplot

      \addplot
      table[x=nodes,y=time (ms),col sep=semicolon] {plots.r1/gnp-0.1-jgrapht-sparse-prim.csv}; 
      \addplot
      table[x=nodes,y=time (ms),col sep=semicolon] {plots.r1/gnp-0.1-jgrapht-sparse-kruskal.csv};
      \addplot
      table[x=nodes,y=time (ms),col sep=semicolon] {plots.r1/gnp-0.1-jgrapht-sparse-boruvka.csv}; 

      \nextgroupplot

      \addplot
      table[x=nodes,y=time (ms),col sep=semicolon] {plots.r1/snap-undirected-jgrapht-sparse-prim.csv}; 
      \addplot
      table[x=nodes,y=time (ms),col sep=semicolon] {plots.r1/snap-undirected-jgrapht-sparse-kruskal.csv};
      \addplot
      table[x=nodes,y=time (ms),col sep=semicolon] {plots.r1/snap-undirected-jgrapht-sparse-boruvka.csv}; 

      \nextgroupplot

      \addplot
      table[x=nodes,y=time (ms),col sep=semicolon] {plots.r1/USA-jgrapht-sparse-prim.csv}; 
      \addplot
      table[x=nodes,y=time (ms),col sep=semicolon] {plots.r1/USA-jgrapht-sparse-kruskal.csv};
      \addplot
      table[x=nodes,y=time (ms),col sep=semicolon] {plots.r1/USA-jgrapht-sparse-boruvka.csv}; 

      \legend{Prim, Kruskal, Bor\r{u}vka}
    \end{groupplot}
  \end{tikzpicture}
  \caption{Execution time of the Prim, Kruskal and Bor\r{u}vka minimum spanning tree algorithms
      using the \jgrapht{} library (sparse backend) with
      Barabasi-Albert (top-left), 
      $Gnp$ with $p=0.1$ (top-right),
      undirected R-MAT ($a=0.57, b=0.19, c=0.19$) (bottom-left), and
      USA roadmaps (bottom-right).
      }\label{elapsed:mst}
\end{figure}
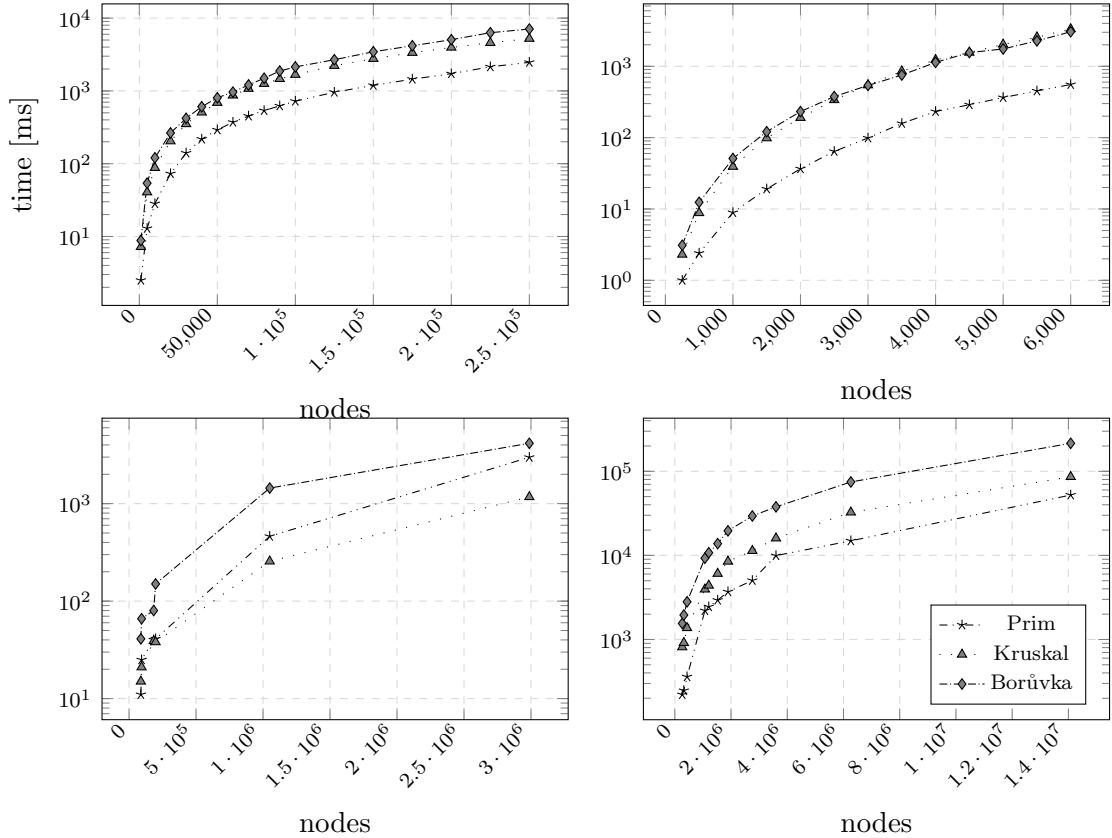

\subsection{Computational Results - Internal comparison}
For many graph problems, \jgrapht{} provides several alternative algorithms which implement a common Java
interface. As these algorithms utilize different underlying techniques, they often exhibit different
runtime-characteristics. Due to their common interface, a user can straightforwardly interchange different
implementations without the need to change code. Since selecting the best algorithm for a given problem
under specific circumstances is not straightforward, in this section we include an internal comparison of
different implementations for two common graph problems, the Minimum Spanning Tree (MST) and the
Maximum-Flow (MF) problem.

\subsubsection{Minimum Spanning Tree} \label{subsub:mst}

\jgrapht{} contains three classic algorithms for solving the MST Problem in weighted, undirected graphs: 
Prim's algorithm ($\Oh(m+n \log n)$), Kruskal's algorithm ($\Oh(m \log n)$) and Bor\r{u}vka's algorithm ($\Oh((m+n)\log n)$). 
The implementation of Prim's algorithm relies on a Fibonacci Heap whereas the other two rely on Union-Find data structures
with the union-by-rank and path-compression heuristics.

For the Barabasi-Albert, $Gnp$ and USA roadmaps graphs, Figure~\ref{elapsed:mst} shows that Prim's algorithm outperforms
Kruskal's algorithm which in turn outperforms Bor\r{u}vka's algorithm. These results are consistent with the results
reported in~\cite{BAHI01}. Here, Prim is 2-3 times faster than the other 2 algorithms. Interestingly, for the R-MAT graphs,
Kruskal's algorithm is roughly 3.8 times faster than Prim's algorithm. In our implementation, Prim's algorithm typically
performs well on denser graphs; on really sparse graphs, Kruskal becomes competitive due to its simplicity
(simpler data-structures). 
Bor\r{u}vka's algorithm is consistently slower than the other 2 algorithms in all experiments. However, its main 
idea (repeated rounds of contractions) has been successfully applied in parallel implementations~\cite{chung1996parallel}. 
Thus, for completeness as well as for research and comparison purposes, we retain all three algorithms as 
we believe that all these techniques should be present in the library.
Ultimately, the end-user decides which algorithm
is best suited for his or her application.

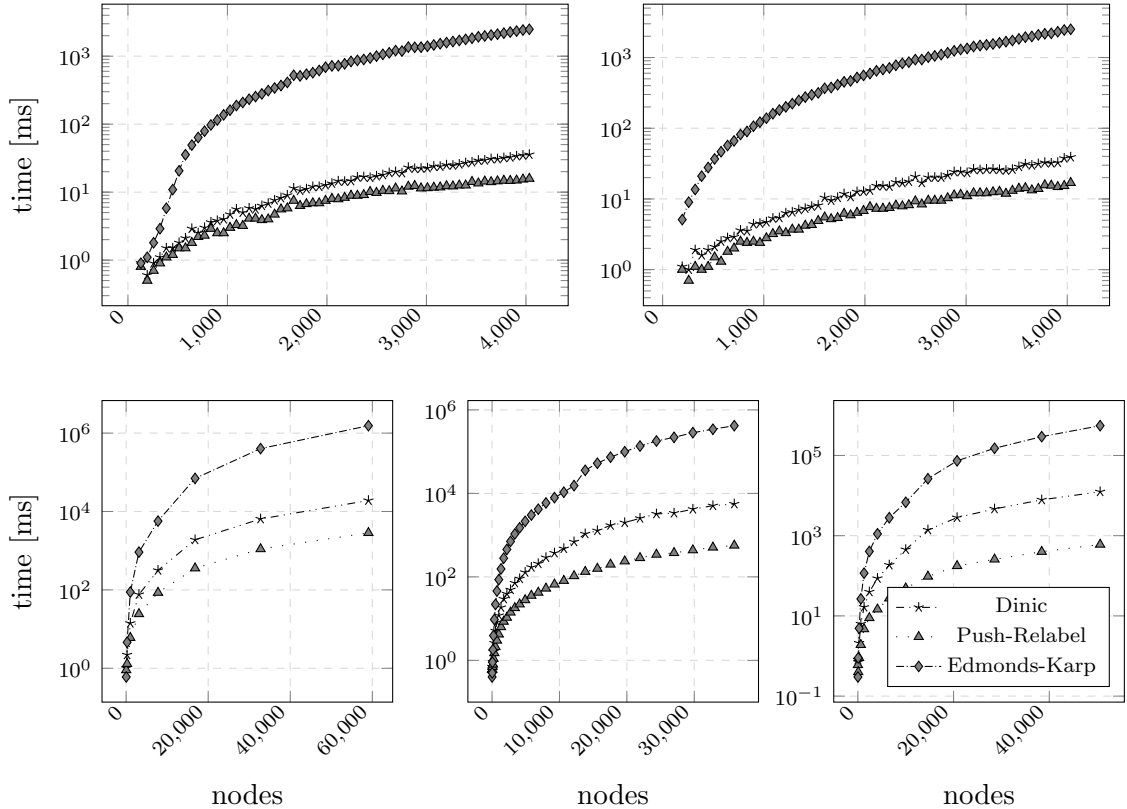
\begin{figure}[t!]
  \begin{tikzpicture}
    
    \begin{groupplot}[
          group style={group size=2 by 1},
          height=0.25\textheight,
          width=0.5\textwidth, 
          grid=major, 
          grid style={dashed,gray!30}, 
          xticklabel style={rotate=45,anchor=east,font=\scriptsize},
          yticklabel style={font=\scriptsize},
          scaled x ticks=false,
          cycle list name=mybw,          
          ymode=log,
          legend columns=1,
          legend style={
            font=\scriptsize,
            at={(0.95,0.05)},
            anchor=south east,
            /tikz/column 2/.style={column sep=1pt},
          }
      ]
      \nextgroupplot[
        ylabel=time,
        y unit=ms,
      ]
      \addplot
          table[x=nodes,y=time (ms),col sep=semicolon] {plots.r1/wide-wash-jgrapht-dinic.csv}; 
      \addplot
          table[x=nodes,y=time (ms),col sep=semicolon] {plots.r1/wide-wash-jgrapht-push-relabel.csv};
      \addplot
          table[x=nodes,y=time (ms),col sep=semicolon] {plots.r1/wide-wash-jgrapht-edmonds-karp.csv};
      \nextgroupplot

      \addplot 
      table[x=nodes,y=time (ms),col sep=semicolon] {plots.r1/long-wash-jgrapht-dinic.csv};
      \addplot
      table[x=nodes,y=time (ms),col sep=semicolon] {plots.r1/long-wash-jgrapht-push-relabel.csv};
      \addplot
      table[x=nodes,y=time (ms),col sep=semicolon] {plots.r1/long-wash-jgrapht-edmonds-karp.csv};

    \end{groupplot}
  \end{tikzpicture}
  \begin{tikzpicture}
    
    \begin{groupplot}[
          group style={group size=3 by 1},
          height=0.25\textheight,
          width=0.35\textwidth, 
          grid=major, 
          grid style={dashed,gray!30}, 
          xlabel=nodes,
          xticklabel style={rotate=45,anchor=east,font=\scriptsize},
          yticklabel style={font=\scriptsize},
          scaled x ticks=false,
          cycle list name=mybw,          
          ymode=log,
          legend columns=1,
          legend style={
            font=\scriptsize,
            at={(0.95,0.05)},
            anchor=south east,
            /tikz/column 2/.style={column sep=1pt},
          }
      ]
      \nextgroupplot[
        ylabel=time,
        y unit=ms,
      ]
      \addplot
          table[x=nodes,y=time (ms),col sep=semicolon] {plots.r1/flat-genrmf-jgrapht-dinic.csv}; 
          \addplot
          table[x=nodes,y=time (ms),col sep=semicolon] {plots.r1/flat-genrmf-jgrapht-push-relabel.csv};
          \addplot
          table[x=nodes,y=time (ms),col sep=semicolon] {plots.r1/flat-genrmf-jgrapht-edmonds-karp.csv};
      \nextgroupplot

      \addplot 
          table[x=nodes,y=time (ms),col sep=semicolon] {plots.r1/square-genrmf-jgrapht-dinic.csv};
          \addplot
          table[x=nodes,y=time (ms),col sep=semicolon] {plots.r1/square-genrmf-jgrapht-push-relabel.csv};
          \addplot
          table[x=nodes,y=time (ms),col sep=semicolon] {plots.r1/square-genrmf-jgrapht-edmonds-karp.csv};

      \nextgroupplot      

      \addplot 
      table[x=nodes,y=time (ms),col sep=semicolon] {plots.r1/long-genrmf-jgrapht-dinic.csv};
      \addplot
      table[x=nodes,y=time (ms),col sep=semicolon] {plots.r1/long-genrmf-jgrapht-push-relabel.csv};
      \addplot
      table[x=nodes,y=time (ms),col sep=semicolon] {plots.r1/long-genrmf-jgrapht-edmonds-karp.csv};

      \legend{Dinic, Push-Relabel, Edmonds-Karp}
    \end{groupplot}
  \end{tikzpicture}
  \caption{
    Execution time of Dinic, Push-Relabel and Edmonds-Karp max-flow algorithms using the \jgrapht{} library
    (default backend).
    Graphs are {\em wide} wash (upper-left), {\em long} wash (upper-right), {\em flat} genrmf (bottom-left), 
      {\em square} genrmf (bottom-middle) and {\em long} genrmf (bottom-right).
    }
  \label{elapsed:max-flow}
\end{figure}

\subsubsection{Maximum Flow} \label{subsub:mf}
\jgrapht{} provides three algorithms to compute Maximum Flows in weighted graphs: Edmonds-Karp algorithm ($\Oh(n m^2)$),
Dinic's algorithm ($\Oh(n^2 m)$) and the Push-Relabel algorithm ($\Oh(n^3)$). For this experiment we use the same experimental set up as in~\cite{flowthesis}. Instead of using the general Barabasi-Albert or $Gnp$ models to generate the instances, we used two dedicated generators\footnote{\url{http://archive.dimacs.rutgers.edu/pub/netflow/generators/}} from the first DIMACS~\cite{johnson1993network} challenge: \textit{RMFGEN}~\cite{goldfarb1988computational} and \textit{washington}.

RMFGEN takes 4 parameters $a$, $b$, $cmin$ and $cmax$ and produces a graph which consists of $b$ layers, each having $a \times a$ nodes laid
out in a square grid. A node in a layer has edges to its adjacent grid nodes, as well as one additional edge to a random node in the next layer.
The resulting graph has $n = a^2 b$ nodes and $m = 4 a (a-1) b + a(b-1)$ edges. The source node is part of the first layer, while the target node
is located in the last layer. Capacities between layers are randomly generated in the range $[cmin, cmax]$; capacities inside layers are big
enough so that all flow can be pushed around inside the layer. Similar to \cite{flowthesis}, we generate three types of graphs: (a) {\em long}
where $a^2=b$, (b) {\em flat} where $a=b^2$, and (c) {\em wide} where $a=b$.

Analogous to RMFGEN, the washington generator is used to produce {\em random level graphs} in which nodes are laid out in rows and columns. Each node is connected to 3 randomly selected nodes in the next column. The source node has edges to all nodes in the first row, whereas the target node is connected to all nodes in the last row. Two types of graphs are generated: (a) {\em wide} graphs having 64 rows and a variable number of columns, and (b) {\em long} graphs with 64 columns and a variable number of rows.

The computational results for each of the 5 graph types are depicted in Figure~\ref{elapsed:max-flow}. In each of the graphs, Push-Relabel has the best performance, followed by Dinic. These algorithms are approximately 489.7 and 58.9 times faster than the well-known Edmonds-Karp maximum flow algorithm. Again these results match the findings reported in~\cite{flowthesis}.

\subsection{Computational Results - Graph Representations}

In this subsection we compare the performance and memory utilization of the different graph representations
(see Section~\ref{subsec:representations}):
(a) the default \jgrapht{} backend,
(b) the default \jgrapht{} backend using the \textit{fastutil}\footnote{\url{http://fastutil.di.unimi.it/}} (v8.2.2) library
for all hashtables, 
(c) a graph adapter which wraps the \guava{}\footnote{\url{https://github.com/google/guava}} (v26.0)
library graph data-structures, and
(d) the sparse backend. 
In particular, for the \guava{} library, we selected two of their graph representations: (a)
\textit{Network}, and (b) \textit{ValueGraph}. The third \guava{} representation, \textit{Graph}, behaves almost identical
to the \textit{ValueGraph} and has therefore been omitted from our figures. 

To measure the computational performance of the different graph representations, we created a portfolio of algorithms, consisting
of Prim's MST algorithm, Dijkstra's shortest path and Edmonds' Maximum Cardinality Matching Algorithm. In the experimental
evaluation, we measure the total time it takes to execute these algorithms on a number of graphs with different representations.
Figure~\ref{elapsed:backend} contains the results of this comparison. The experiments are conducted using the exact same
settings as in the previous experiments. As can be observed from Figure~\ref{elapsed:backend}, the default, sparse and
fastutil \jgrapht{} representations have rather comparable performance profiles: the sparse backend is 1.2, resp. 1.4 times
faster than the default or the one with fastutil.
In contrast, both Guava representations are significantly slower, despite the fact that the Guava adapter classes in
\jgrapht{} merely translate calls to the underlying Guava implementations, and thus do not impact the performance significantly.
For reference, Guava ValueGraph and Network graph representations are resp.~4.1 and 3 times slower than \jgrapht{}'s sparse backend.

\begin{figure}[t!]
  \begin{tikzpicture}
    
    \begin{groupplot}[
          group style={
            group size=2 by 2,
            vertical sep=1.5cm,
          },
          height=0.25\textheight,
          width=0.50\textwidth,       
          grid=major, 
          grid style={dashed,gray!30}, 
          xlabel=nodes,
          xticklabel style={rotate=45,anchor=east,font=\scriptsize},
          yticklabel style={font=\scriptsize},
          scaled x ticks=false,
          cycle list name=mybw,
          legend columns=3,
          legend style={
            at={(1.0,2.5)},
            anchor=south east,
            /tikz/column 2/.style={column sep=5pt},
          }
      ]
      \nextgroupplot[
        ylabel=bytes/edge,
      ]

      \addplot
      table[x=nodes,y=peredge,col sep=semicolon] {plots.r1/barabasi-albert-20-10-jgrapht-memory.csv};

      \addplot
      table[x=nodes,y=peredge,col sep=semicolon] {plots.r1/barabasi-albert-20-10-jgrapht-sparse-memory.csv};

      \addplot
      table[x=nodes,y=peredge,col sep=semicolon] {plots.r1/barabasi-albert-20-10-fastutil-memory.csv};

      \addplot
      table[x=nodes,y=peredge,col sep=semicolon] {plots.r1/barabasi-albert-20-10-guava-network-memory.csv};

      \addplot
      table[x=nodes,y=peredge,col sep=semicolon] {plots.r1/barabasi-albert-20-10-guava-valuegraph-memory.csv};

      \nextgroupplot
      \addplot
      table[x=nodes,y=peredge,col sep=semicolon] {plots.r1/gnp-0.1-jgrapht-memory.csv};

      \addplot
      table[x=nodes,y=peredge,col sep=semicolon] {plots.r1/gnp-0.1-jgrapht-sparse-memory.csv};

      \addplot
      table[x=nodes,y=peredge,col sep=semicolon] {plots.r1/gnp-0.1-fastutil-memory.csv};

      \addplot
      table[x=nodes,y=peredge,col sep=semicolon] {plots.r1/gnp-0.1-guava-network-memory.csv};

      \addplot
      table[x=nodes,y=peredge,col sep=semicolon] {plots.r1/gnp-0.1-guava-valuegraph-memory.csv};

    \nextgroupplot[
      xmode=log,
      ylabel=bytes/edge,
    ]

    \addplot
    table[x=nodes,y=peredge,col sep=semicolon] {plots.r1/rmat-undirected-jgrapht-memory.csv};

    \addplot
    table[x=nodes,y=peredge,col sep=semicolon] {plots.r1/rmat-undirected-jgrapht-sparse-memory.csv};    

    \addplot
    table[x=nodes,y=peredge,col sep=semicolon] {plots.r1/rmat-undirected-fastutil-memory.csv};

    \addplot
    table[x=nodes,y=peredge,col sep=semicolon] {plots.r1/rmat-undirected-guava-network-memory.csv};

    \addplot
    table[x=nodes,y=peredge,col sep=semicolon] {plots.r1/rmat-undirected-guava-valuegraph-memory.csv};

      \nextgroupplot[
        xlabel=edges,
        xmode=log,
      ]
      \addplot
      table[x=nodes,y=peredge,col sep=semicolon] {plots.r1/snap-undirected-jgrapht-memory.csv};
      
      \addplot
      table[x=nodes,y=peredge,col sep=semicolon] {plots.r1/snap-undirected-jgrapht-sparse-memory.csv};

      \addplot
      table[x=nodes,y=peredge,col sep=semicolon] {plots.r1/snap-undirected-fastutil-memory.csv};
      
      \addplot
      table[x=nodes,y=peredge,col sep=semicolon] {plots.r1/snap-undirected-guava-network-memory.csv};

      \addplot
      table[x=nodes,y=peredge,col sep=semicolon] {plots.r1/snap-undirected-guava-valuegraph-memory.csv};
 
      \legend{\jgrapht{}, \jgraphtS{}, \fastutil{}, \guavaN{}, \guavaV{}}
               
    \end{groupplot}
  \end{tikzpicture}
  \caption{Space requirements for different graph representations
  with Barabasi-Albert (up-left), 
   $Gnp$ with $p=0.1$ (up-right), 
   undirected R-MAT ($a=0.57, b=0.19, c=0.19$) (bottom-left), 
   and SNAP undirected graphs (bottom-right).
  }\label{space:backend}
\end{figure}
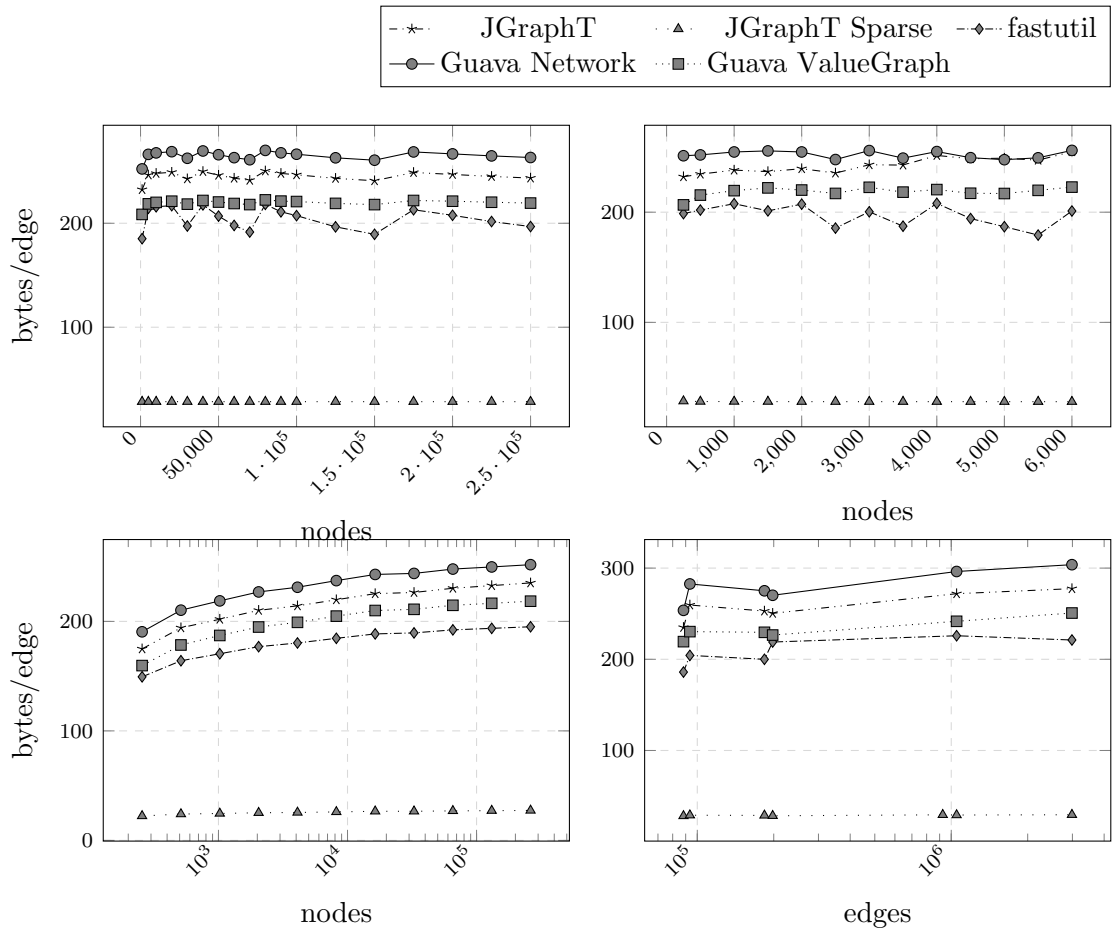

Finally, a comparison of the memory-footprint of the various graph representations is presented in Figure~\ref{space:backend}.
In this particular experiment, we measure the memory utilization of different representations for different types of graphs inside the JVM.
When the graph representation has no native support for edge weights, we
wrapped the graph inside \jgrapht{}'s \jmint{AsWeightedGraph} adapter class. Performing memory measurements in the
JVM is a somewhat involved process for which we used the specialized Jamm software package~\cite{jbellis}. In short,
Jamm loads a Java agent which internally relies on the \verb|Instrumentation.getObjectSize| method from the
\verb|java.lang.instrument| package to measure the amount of space occupied by Java objects.
In order to present the final results we normalize the space utilization per graph by dividing by the total number of edges in
the graph. 

When comparing the results in Figure~\ref{space:backend}, it is obvious that \jgrapht{}'s \textit{sparse}
representation has the smallest memory footprint, followed by the \textit{fastutil} and Guava's ValueGraph. The latter two require
resp.~7.1 and 7.8 times more memory than the sparse representation. These results are consistent across the different graphs.
\jgrapht{}'s default representation, as well as Guava's Network graph are the most memory intensive (resp.~8.6 and 9.3 time more
memory than the sparse one). For the largest $Gnp$ graphs, both representations require the same amount of memory. Overall, when comparing
both space and computational efficiency, \jgrapht{}'s sparse representation yields the best performance characteristics. The use of fastutil hashtables
 performs slightly slower than the default representation, but requires significantly less memory. Finally, the Guava representation
should only be used when interoperability with Guava is required.

\begin{figure}[t!]
  \begin{tikzpicture}
    
    \begin{groupplot}[
          group style={
            group size=2 by 2,
            vertical sep=1.5cm,
          },
          height=0.25\textheight,
          width=0.50\textwidth, 
          grid=major, 
          grid style={dashed,gray!30}, 
          xlabel=nodes,
          xticklabel style={rotate=45,anchor=east,font=\scriptsize},
          yticklabel style={font=\scriptsize},
          scaled x ticks=false,
          cycle list name=mybw,
          legend columns=3,
          legend style={
            at={(1.0,2.6)},
            anchor=south east,
            /tikz/column 2/.style={column sep=5pt},
          }
      ]
      \nextgroupplot[
        ylabel=time,
        y unit=ms,
        ymode=log,
      ]

      \addplot
      table[x=nodes,y=time (ms),col sep=semicolon] {plots.r1/barabasi-albert-20-10-jgrapht-bundle.csv};

      \addplot
      table[x=nodes,y=time (ms),col sep=semicolon] {plots.r1/barabasi-albert-20-10-jgrapht-sparse-bundle.csv};
      
      \addplot
      table[x=nodes,y=time (ms),col sep=semicolon] {plots.r1/barabasi-albert-20-10-fastutil-bundle.csv};

      \addplot
      table[x=nodes,y=time (ms),col sep=semicolon] {plots.r1/barabasi-albert-20-10-guava-bundle.csv};

      \addplot
      table[x=nodes,y=time (ms),col sep=semicolon] {plots.r1/barabasi-albert-20-10-guava-valuegraph-bundle.csv};

      \nextgroupplot[
        ymode=log,
      ]
      \addplot
      table[x=nodes,y=time (ms),col sep=semicolon] {plots.r1/gnp-0.1-jgrapht-bundle.csv};

      \addplot
      table[x=nodes,y=time (ms),col sep=semicolon] {plots.r1/gnp-0.1-jgrapht-sparse-bundle.csv};

      \addplot
      table[x=nodes,y=time (ms),col sep=semicolon] {plots.r1/gnp-0.1-fastutil-bundle.csv};

      \addplot
      table[x=nodes,y=time (ms),col sep=semicolon] {plots.r1/gnp-0.1-guava-bundle.csv};

      \addplot
      table[x=nodes,y=time (ms),col sep=semicolon] {plots.r1/gnp-0.1-guava-valuegraph-bundle.csv};

      \nextgroupplot[
        ylabel=time,
        y unit=ms,
        xmode=log,
        ymode=log,
      ]

      \addplot
      table[x=nodes,y=time (ms),col sep=semicolon] {plots.r1/rmat-undirected-jgrapht-bundle.csv};
      
      \addplot
      table[x=nodes,y=time (ms),col sep=semicolon] {plots.r1/rmat-undirected-jgrapht-sparse-bundle.csv};

      \addplot
      table[x=nodes,y=time (ms),col sep=semicolon] {plots.r1/rmat-undirected-fastutil-bundle.csv};

      \addplot
      table[x=nodes,y=time (ms),col sep=semicolon] {plots.r1/rmat-undirected-guava-bundle.csv};

      \addplot
      table[x=nodes,y=time (ms),col sep=semicolon] {plots.r1/rmat-undirected-guava-valuegraph-bundle.csv};

      \nextgroupplot[
        xlabel=edges,
        xmode=log,
        ymode=log,
      ]
      \addplot
      table[x=nodes,y=time (ms),col sep=semicolon] {plots.r1/snap-undirected-jgrapht-bundle.csv};
      
      \addplot
      table[x=nodes,y=time (ms),col sep=semicolon] {plots.r1/snap-undirected-jgrapht-sparse-bundle.csv};

      \addplot
      table[x=nodes,y=time (ms),col sep=semicolon] {plots.r1/snap-undirected-fastutil-bundle.csv};

      \addplot
      table[x=nodes,y=time (ms),col sep=semicolon] {plots.r1/snap-undirected-guava-bundle.csv};

      \addplot
      table[x=nodes,y=time (ms),col sep=semicolon] {plots.r1/snap-undirected-guava-valuegraph-bundle.csv};

      \legend{\jgrapht{}, \jgraphtS{}, \fastutil{}, \guavaN{}, \guavaV{}}

    \end{groupplot}
  \end{tikzpicture}
  \caption{Performance comparison for different graph representations with Barabasi-Albert
  (up-left), 
   $Gnp$ with $p=0.1$ (up-right), 
   undirected R-MAT ($a=0.57, b=0.19, c=0.19$) (bottom-left), 
   and SNAP undirected graphs (bottom-right).
  }\label{elapsed:backend}
\end{figure}
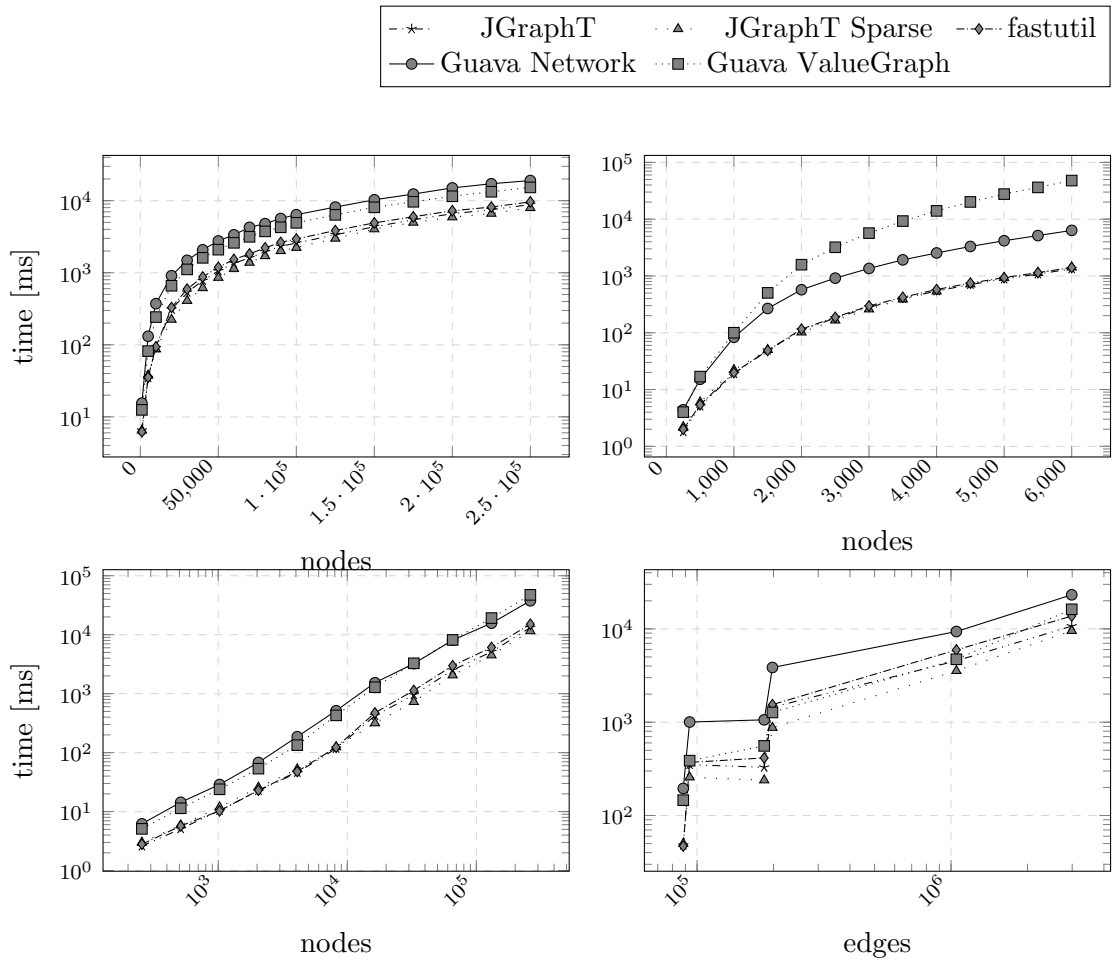

\section{Conclusions}\label{sec:conclusions}

We have presented in detail the motivation, development choices, features, algorithmic support and
internals of the \jgrapht{} library. The library has been in development for more than a decade and
is currently deployed in a variety of commercial products, as well as academic research projects. A major challenge
in the design of a graph library is the delicate balance between expressiveness and performance. In \jgrapht{},
vertices and edges can be represented by any type of object, thereby enabling the user to represent virtually
any type of network. 

In the experimental evaluation, we compared the performance and memory utilization of \jgrapht{} across a number
of alternative open-source graph libraries using a diverse set of input graphs ranging from USA roadmaps and real-world
instances acquired from SNAP, up to generated instances using well-known random generators such as R-MAT,
Barabasi-Albert and Gnp.
The experimental evaluation shows that \jgrapht{} is highly competitive both performance-wise and feature-wise 
since it includes a large and diverse set of graph algorithms coupled together with efficient graph
representations.

\section*{Acknowledgments}
The authors of this paper are the current and past administrators, as well as the main developers and 
maintainers of \jgrapht{}. Collectively, the authors are responsible for the design and development of
the library, scientific soundness, quality assurance of code and documentation, and release management.
Clearly, a project like \jgrapht{} could not exist without contributions from independent contributors.
We would like to thank all contributors to the project and especially Timofey Chudakov, Semen Chudakov,
Alexandru V{\u a}leanu, Ilya Razenshteyn, Alexey Kudinkin, and Philipp Kaesgen for their effort in
improving the library. A complete list of all contributors, as well as their contributions, can be
found at the project's GitHub\footnote{\url{https://github.com/jgrapht/jgrapht}} page.


\appendix

\section{Real-world Datasets}\label{appendix:snap}

The real-world datasets used in our experiments were acquired from SNAP~\cite{leskovec2016snap} 
and can be found in Table~\ref{table:snap:datasets}.

\begin{table}[h]
  \begin{tabular}{l|c|c|c}
     \toprule
     name & type & \#nodes & \#edges\\
     \midrule
     \verb|p2p-Gnutella24| & directed & 26518 & 65369\\
     \verb|wiki-Vote| & directed & 7115 & 103689\\
     \verb|email-EuAll| & directed & 265214 & 420045\\
     \verb|soc-Epinions1| & directed & 75879 & 508837\\
     \verb|ego-Twitter| & directed & 81306 & 1768149\\
     \verb|web-Stanford| & directed & 281903 & 2312497\\
     \verb|web-Google| & directed & 875713 & 5105039\\
     \verb|soc-Pokec| & directed & 1632803 & 30622564\\
     \midrule
     \verb|ego-Facebook| & undirected & 4039 & 88234\\
     \verb|ca-CondMat| & undirected & 23133 & 93497\\
     \verb|email-Enron| & undirected & 36692 & 183831\\
     \verb|ca-AstroPh| & undirected & 18772 & 198110\\
     \verb|com-Amazon| & undirected & 334863 & 925872\\
     \verb|com-DBLP| & undirected & 317080 & 1049866\\
     \bottomrule
  \end{tabular}
  \caption {SNAP dataset.}\label{table:snap:datasets}
  \end{table}

\end{document}